\documentclass[published]{nst}

\usepackage{subfigure,dcolumn}
\usepackage{epstopdf}
\usepackage{mhchem}
\usepackage{upgreek}
\usepackage{amsmath}
\usepackage{cleveref}
\usepackage{siunitx}
\usepackage{verbatim}
\usepackage[section]{placeins}
\crefname{figure}{Fig.}{Fig.}

\begin{document}

\title{Development of the CEPC analog hadron calorimeter prototype}
\thanks{Supported by National Key Program for S\&T Research and Development (2018YFA0404303), the National Natural Science Foundation of China(12125505,11961141006) and Shanghai Pilot Program for Basic Research—Shanghai Jiao Tong University(21TQ1400209).}

\author{Yukun Shi}
\affiliation{State Key Laboratory of Particle Detection and Electronics, University of Science and Technology of China, Hefei 230026, China}
\affiliation{University of Science and Technology of China, Department of Modern Physics, Hefei 230026, China}

\author{Anshun Zhou}
\affiliation{State Key Laboratory of Particle Detection and Electronics, University of Science and Technology of China, Hefei 230026, China}
\affiliation{University of Science and Technology of China, Department of Modern Physics, Hefei 230026, China}

\author{Hao Liu}
\affiliation{State Key Laboratory of Particle Detection and Electronics, University of Science and Technology of China, Hefei 230026, China}
\affiliation{University of Science and Technology of China, Department of Modern Physics, Hefei 230026, China}

\author{Jiechen Jiang}
\affiliation{Institute of High Energy Physics, Chinese Academy of Sciences (CAS), 100049, Beijing, China}

\author{Yanyun Duan}
\affiliation{Tsung-Dao Lee Institute, Shanghai Jiao Tong University, Shanghai, 201210, China}
\affiliation{Key Laboratory for Particle Physics,Astrophysics and Cosmology (Ministry of Education), Shanghai Key Laboratory for Particle Physics and Cosmology,Shanghai 200240,China}

\author{Yunlong Zhang}
\email[Corresponding author, ]{ylzhang@ustc.edu.cn}
\affiliation{State Key Laboratory of Particle Detection and Electronics, University of Science and Technology of China, Hefei 230026, China}
\affiliation{University of Science and Technology of China, Department of Modern Physics, Hefei 230026, China}

\author{Zhongtao Shen}
\affiliation{State Key Laboratory of Particle Detection and Electronics, University of Science and Technology of China, Hefei 230026, China}
\affiliation{University of Science and Technology of China, Department of Modern Physics, Hefei 230026, China}

\author{Jianbei Liu}
\affiliation{State Key Laboratory of Particle Detection and Electronics, University of Science and Technology of China, Hefei 230026, China}
\affiliation{University of Science and Technology of China, Department of Modern Physics, Hefei 230026, China}

\author{Boxiang Yu}
\affiliation{State Key Laboratory of Particle Detection and Electronics, Institute of High Energy Physics, 100049, Beijing, China}
\affiliation{Institute of High Energy Physics, Chinese Academy of Sciences (CAS), 100049, Beijing, China}
\affiliation{School of Physical Science, University of Chinese Academy of Sciences, 100049, Beijing, China}

\author{Shu Li}
\affiliation{{Tsung-Dao Lee Institute \& School of Physics and Astronomy, Shanghai Jiao Tong University, Shanghai 201210, China}}
\affiliation{Key Laboratory for Particle Physics,Astrophysics and Cosmology (Ministry of Education), Shanghai Key Laboratory for Particle Physics and Cosmology,Shanghai 200240,China}

\author{Haijun Yang}
\affiliation{{School of Physics and Astronomy \& Tsung-Dao Lee Institute, Shanghai Jiao Tong University, Shanghai 200240, China}}
\affiliation{Key Laboratory for Particle Physics,Astrophysics and Cosmology (Ministry of Education), Shanghai Key Laboratory for Particle Physics and Cosmology,Shanghai 200240,China}

\author{Yong Liu}
\affiliation{State Key Laboratory of Particle Detection and Electronics, Institute of High Energy Physics, 100049, Beijing, China}
\affiliation{Institute of High Energy Physics, Chinese Academy of Sciences (CAS), 100049, Beijing, China}

\author{Liang Li}
\affiliation{Institute of High Energy Physics, Chinese Academy of Sciences (CAS), 100049, Beijing, China}
\affiliation{Department of Nuclear Technology and Application,China Institute of Atomic Energy,102413,Beijing,China}

\author{Zhen Wang}
\affiliation{Tsung-Dao Lee Institute, Shanghai Jiao Tong University, Shanghai, 201210, China}
\affiliation{Key Laboratory for Particle Physics,Astrophysics and Cosmology (Ministry of Education), Shanghai Key Laboratory for Particle Physics and Cosmology,Shanghai 200240,China}

\author{Siyuan Song}
\affiliation{{School of Physics and Astronomy \& Tsung-Dao Lee Institute, Shanghai Jiao Tong University, Shanghai 200240, China}}
\affiliation{Key Laboratory for Particle Physics,Astrophysics and Cosmology (Ministry of Education), Shanghai Key Laboratory for Particle Physics and Cosmology,Shanghai 200240,China}

\author{Dejing Du}
\affiliation{Institute of High Energy Physics, Chinese Academy of Sciences (CAS), 100049, Beijing, China}

\author{Jiaxuan Wang}
\affiliation{State Key Laboratory of Particle Detection and Electronics, University of Science and Technology of China, Hefei 230026, China}
\affiliation{University of Science and Technology of China, Department of Modern Physics, Hefei 230026, China}

\author{Junsong Zhang}
\affiliation{Institute of High Energy Physics, Chinese Academy of Sciences (CAS), 100049, Beijing, China}

\author{Quan Ji}
\affiliation{Institute of High Energy Physics, Chinese Academy of Sciences (CAS), 100049, Beijing, China}

\begin{abstract}
	The Circular Electron Positron Collider (CEPC) is a next-generation electron–positron collider proposed for the precise measurement of the properties of the Higgs boson. To emphasize boson separation and jet reconstruction, the baseline design of the CEPC detector was guided by the particle flow algorithm (PFA) concept. As one of the calorimeter options, the analogue hadron calorimeter (AHCAL) was proposed. The CEPC AHCAL comprises a 40-layer sandwich structure using steel plates as absorbers and scintillator tiles coupled with silicon photomultipliers (SiPM) as sensitive units. \textcolor{black}{To validate the feasibility of the AHCAL option, a series of studies were conducted to develop a prototype. This AHCAL prototype underwent an electronic test and a cosmic ray test to assess its performance and ensure it was ready for three beam tests performed in 2022 and 2023. The test beam data is currently under analysis, and the results are expected to deepen our understanding of hadron showers, validate the concept of Particle Flow Algorithm (PFA), and ultimately refine the design of the CEPC detector.}
  
\end{abstract}


\maketitle

\section{Introduction}
	The discovery of the Higgs boson marked a significant milestone in the field of particle physics \cite{atlas2012observation,cms2012observation}. Now, the focus is on precisely measuring the properties of the Higgs boson\cite{Higgs1,Higgs2,Higgs3,Higgs4}, which has led to the proposal of the Circular Electron-Positron Collider (CEPC)  as a future Higgs factory\cite{CEPC_CDR1,CEPC_CDR2}.  The CEPC could operate at a center-of-mass energy $\sqrt{s} \sim$\SI{240}{GeV} with a luminosity of $8.3 \times 10^{34}$\ \si{cm^{-2}s^{-1}}, resulting in an integrated luminosity of \SI{21.6}{ab^{-1}} for two interaction points over a decade, producing 4 million Higgs bosons\cite{CEPC_TDR,Yang:2023iik}. In addition, it will also be operated on the $Z$-pole as a $Z$ factory, perform a precise $WW$ threshold scan, and be upgraded to a center-of-mass energy of \SI{360}{GeV}, close to the $t\overline{t}$ threshold.

    The CEPC physics potential has been continuously explored by the CEPC Physics study groups, focusing on a wide range of topics, including Higgs precision measurements, precise EW measurements, Flavor Physics, and so on. Many Higgs boson couplings can be measured with precision about one order of magnitude better than those achievable at the High Luminosity LHC (HL-LHC)\cite{CEPC_Higgs_precision,CEPC_Higgs_CrossSection,CEPC_Higgs_BR,CEPC_Higgs_ZZ,tan2020search}. In addition, the CEPC is expected to improve the current precision of many of the electroweak observables by about one order of magnitude or more\cite{CEPC_CDR2,CEPC_EW}. 
    
\begin{figure}[htbp]
        \centering    
        \includegraphics[width=.35\textwidth]{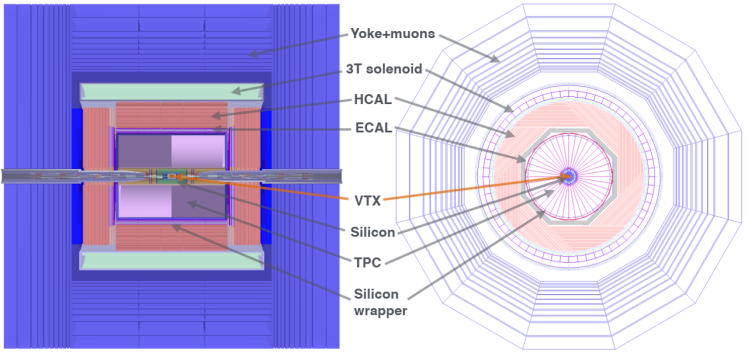}
        \caption{The overview of the CEPC baseline detector concept. In the barrel from inner to outer, the detector is composed of a silicon pixel vertex detector, a silicon inner tracker, a TPC, a silicon wrapper, an Electromagnetic Calorimeter(ECAL), a Hadronic Calorimeter(HCAL), a solenoid of 3 Tesla and a return yoke with embedded a muon detector.\cite{CEPC_CDR2}.}
        \label{Fig:CEPC_detector}  
    \end{figure}

    These CEPC physics studies also identified a handful of critical detector requirements. A boson mass resolution(BMR) of 4\% is required to separate the Higgs bosons from the W and Z bosons in their hadronic decays, corresponding to an unprecedented jet energy resolution of 3-4\% at \SI{100}{GeV}\cite{tan2020search}.To achieve this jet energy resolution, the baseline detector concept was guided by the particle flow algorithm (PFA) of measuring final state particles in the most suited detector subsystem\cite{thomson2009particle,Liu:2024pah}. It employs an ultra-high granular calorimetry system to efficiently separate the final state particle showers, a low material tracking system to minimize the interaction of the final state particles in the tracking material, and a large volume 3 Tesla solenoid that encloses the entire calorimetry system as illustrated in \cref{Fig:CEPC_detector}\cite{CEPC_CDR2}.

    The calorimetry system plays a crucial role in the PFA for separating different particles in a jet and measuring the photons and neutral hadrons. The analog hadron calorimeter(AHCAL) with ultra-high granularity was proposed as one of the options for the CEPC calorimetry. The AHCAL utilizing the SiPM-on-tile technology was previously developed by the CALICE group for the future linear collider, and a prototype was exposed to test beams in 2018\cite{Sefkow_2018rhp,Laudrain_2022,white2023design}. However, considering that the CEPC will operate at a relatively low center-of-mass energy to prioritize precise Higgs measurements, it is essential to optimize the design of the AHCAL and build a prototype to validate its performance.

    This article describes the design, construction, and tests of the CEPC AHCAL prototype that consists of a 40 layer steel-scintillator sandwich structure. The design of the AHCAL prototype including the sensitive units, the readout electronics and the mechanical structure are introduced in Section 2. Section 3 discusses the construction of the AHCAL prototype, covering the production and testing of scintillator tiles, SiPM testing, sensitive layer production, and prototype integration. In Section 4, the electronic test and the cosmic ray test are presented, demonstrating the functionality of the prototype. Finally, Section 5 provides a summary. 
         
\section{Design of the AHCAL prototype}

\begin{figure}[htbp]
    \centering    
    \includegraphics[width=.35\textwidth]{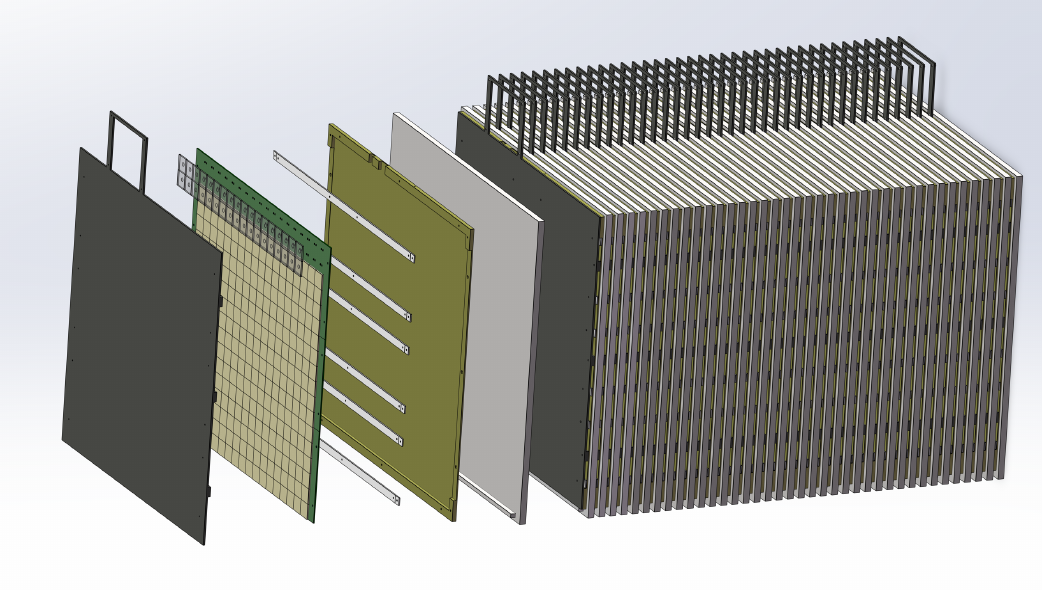}
    \caption{The scintillator-steel sandwich structure of the AHCAL prototype. The scintillator tiles and PCB board were housed within the steel cassette, creating a sensitive layer for the AHCAL.}
    \label{Fig:prototype_structure}  
    \end{figure}

    The design of the CEPC AHCAL was optimized based on the PFA reconstruction results of the $H \rightarrow gg$ channel within the CEPC software environment\cite{Zhao:2018jiq}. The optimized AHCAL design could achieve a Higgs boson mass resolution of 3.7\%\cite{shi2022design}. Following the optimized AHCAL design, the AHCAL prototype was expected to exhibit an energy linearity of approximately 1.5\% and an energy resolution of approximately $\frac{60\%}{\sqrt{E(\si{GeV})}}$ for pion particles with an energy ranging from \SI{10}{GeV} to \SI{80}{GeV}.The prototype consists of 40 sampling layers, with each layer composed of \SI{20}{mm} steel as the absorber and $40\times40\times3$\ \si{mm^3} plastic scintillator tiles as the sensitive material, as illustrated in \cref{Fig:prototype_structure}. The total material of the prototype amounts to approximately \SI{5}{\lambda}. The sensitive area of the prototype is $720 \times 720$\ \si{mm^2}, corresponding to 12960 scintillator tiles. These Scintillator tiles are read out by the silicon photomultiplier (SiPM) and SPIROC2E chips. The scintillator tiles and electronics are housed in steel cassettes forming sensitive layers, which are then secured by the supporting framework along with the absorber plates.

\subsection{Sensitive unit}

\begin{figure}[htbp]
        \centering
        \subfigure[]{
            \includegraphics[width=.35\textwidth]{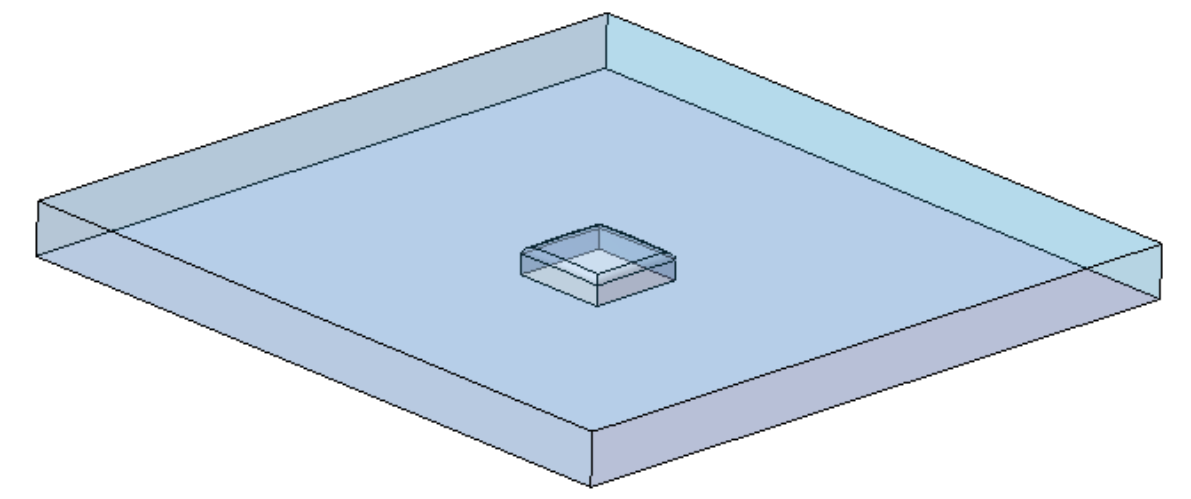}
            \label{Fig:scintillator1} 
        }
        \subfigure[]{
            \includegraphics[width=.35\textwidth]{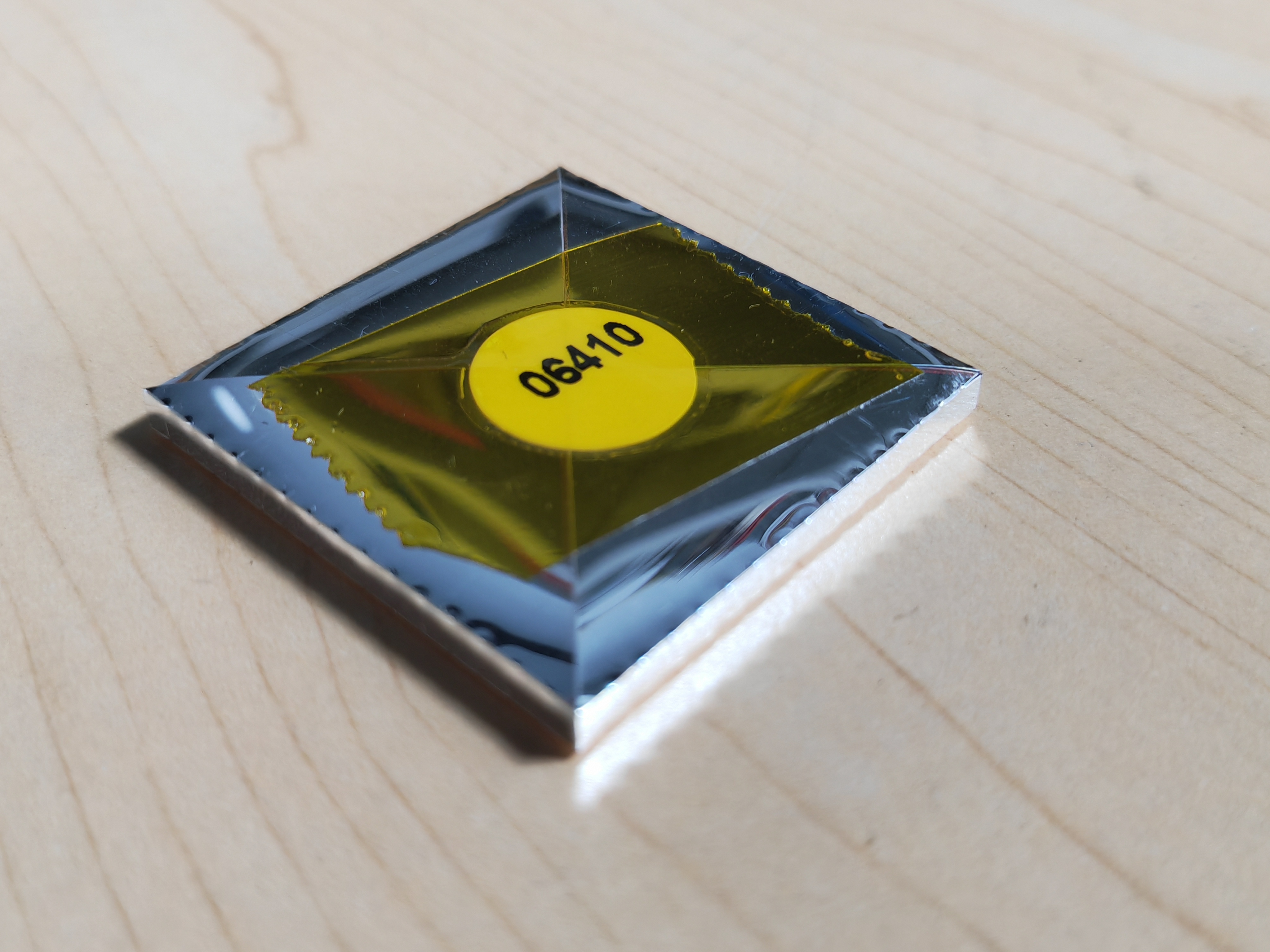}
            \label{Fig:scintillator4} 
        }
        \subfigure[]{
            \includegraphics[width=.35\textwidth]{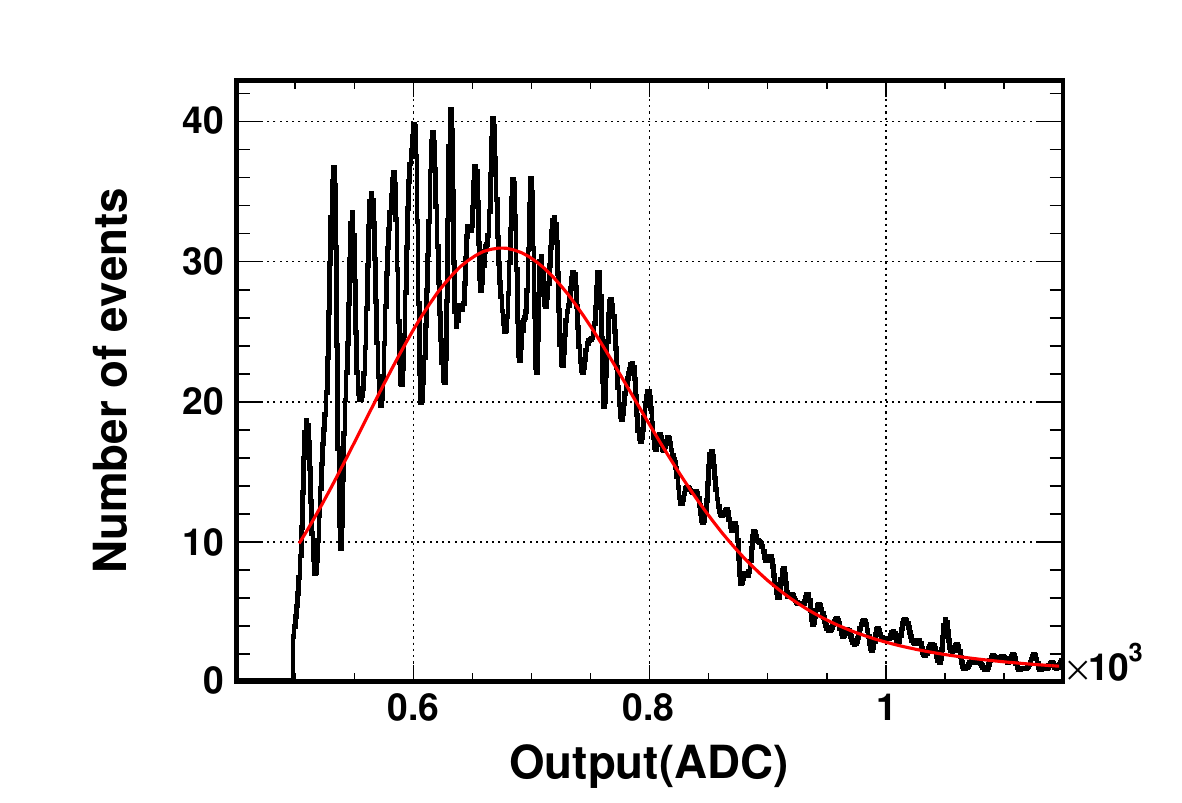}
            \label{Fig:LY_S14}
        }
        \caption{(a) The schematic layout of the AHCAL scintillator tile, with a groove at the bottom to accommodate the SiPM and LED. (b) A scintillator tile wrapped with the ESR film. (c) The scintillator light response to Sr-90, from which the light yield of the scintillator was analyzed.}
          
    \end{figure}

    The scintillator is often used as the sensitive material for sampling calorimetry, along with the novel photosensitive device SiPM\cite{He2023,Jiang_2020,Wu_2018,Wan:2023ssz,Tang:2024,Yang:2024,Liu:2024,Wang:2023,Zhang:2023}. The AHCAL sensitive unit is composed of a scintillator tile and a SiPM. The $40\times40\times3$\ \si{mm^3} scintillator tile is designed with a $5.5\times5.5\times1.1$\ \si{mm^3} groove at the bottom to accommodate the SiPM\cite{Li_2021}, as depicted in \cref{Fig:scintillator1}. Moreover, an LED is also positioned in the groove adjacent to the SiPM for calibration\cite{Jiang_2020}.
    
    The scintillator tiles were produced using a cost-effective injection molding technique, with the recipes and procedures undergoing eight iterations to achieve optimal performance\cite{Jiang_2020}. Subsequently, the scintillator tile was wrapped with ESR, as shown in \cref{Fig:scintillator4}. 
  
    The scintillator tile was tested by a Sr-90 source and a S14160-1315P SiPM\cite{HAMMATSU,Wang:2023ffl}. The result was shown in \cref{Fig:LY_S14}, indicating that the light yield of this scintillator tile was approximately 17 p.e. (photon-electron). The non-uniformity of the scintillator tile was obtained by varying the position of the Sr-90 source. It turned out to be approximately 6.7\%\cite{Li_2021}.

    A simulation based on GEANT4 was carried out to investigate the impact of the SiPM on the performance of the AHCAL. A standalone AHCAL prototype geometry was built, and its responses to hadrons of different energies were simulated.
 
    Various thresholds in the unit of MIP(Minimum Ionizing Particle) were applied in the simulation. The energy resolution improves while lower threshold is utilized, indicating that the SiPM with low crosstalk or high light yield should be selected for the prototype.


    The energy deposition in a single AHCAL scintillator tile for \SI{100}{GeV} hadrons was also simulated\cite{Jiang_SiPM2}. The majority of scintillator tiles exhibit energy depositions below 400 MIP, indicating that the dynamic range of the SiPM, estimated as the pixel number divided by the light yield, should exceed 400.


    To address the requirements of the CEPC AHCAL, several sensitive units of different SiPMs were evaluated\cite{Jiang_SiPM1, Jiang_SiPM2}. \textcolor{black}{As a result, two types of SiPMs were selected for the AHCAL prototype, with their parameters listed in \cref{tab:SiPM}. Note that the EQR-22-1313D-S model has four pads, but only two of them were read out.}
    the S14160-1315P SiPM was implemented on 38 AHCAL sensitive layers due to its low crosstalk and high dynamic-range. The EQR-22-1313D-S SiPM was utilized on the last 2 layers of the AHCAL prototype due to its high light yield and low price\cite{NDL,Liu:2019lgs}.

    \begin{table}[htb]
        \centering
        \caption{Technical parameters of SiPMs implemented in the AHCAL prototype}
        \resizebox{0.5\textwidth}{.1\textwidth}{
        \begin{tabular}{|c|c|c|}
        \hline
        SiPM & HPK S14160-1315PS & NDL EQR15-22-1313D-S \\
        \hline
        Sensitive area(\si{mm^2}) & $1.3\times1.3$  & $4\times1.3\times1.3$ \\
        \hline
        Number of pixels & $7248$ & $7396\times4$ \\
        \hline
        Break down voltage(\si{V})  & 38 & 27.5 \\
        \hline
        Dark count rate(\si{kHz})  & 200 & 400 \\
        \hline
        Photon detection efficiency(PDE)(\%)  & 32 & 45\\
        \hline
        Gain($10^{5}$) & 1.93 & 4.0 \\
        \hline
        Cross talk(\%) & $<1$ & 4.4 \\
        \hline
        light yield(p.e.) & 17 & 40(2 pads) \\
        \hline
        \end{tabular}}
        \label{tab:SiPM}
    \end{table}

\subsection{Readout electronics system}

\begin{figure}[htbp]
    \centering
    \subfigure[]{
        \includegraphics[width=.35\textwidth]{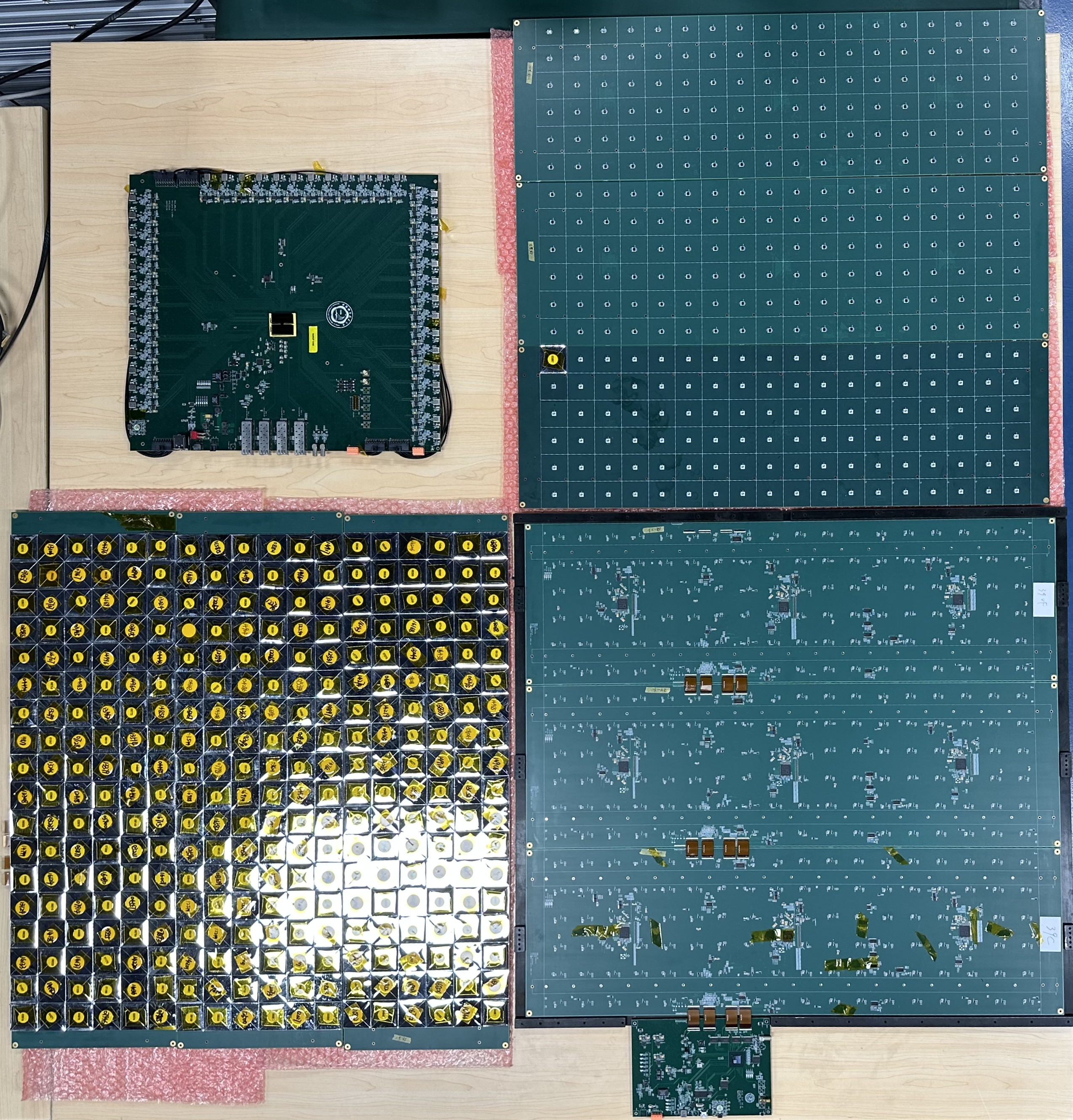}
        \label{Fig:Boards}
    }
    \subfigure[]
    {
            \centering
            \includegraphics[width=.35\textwidth]{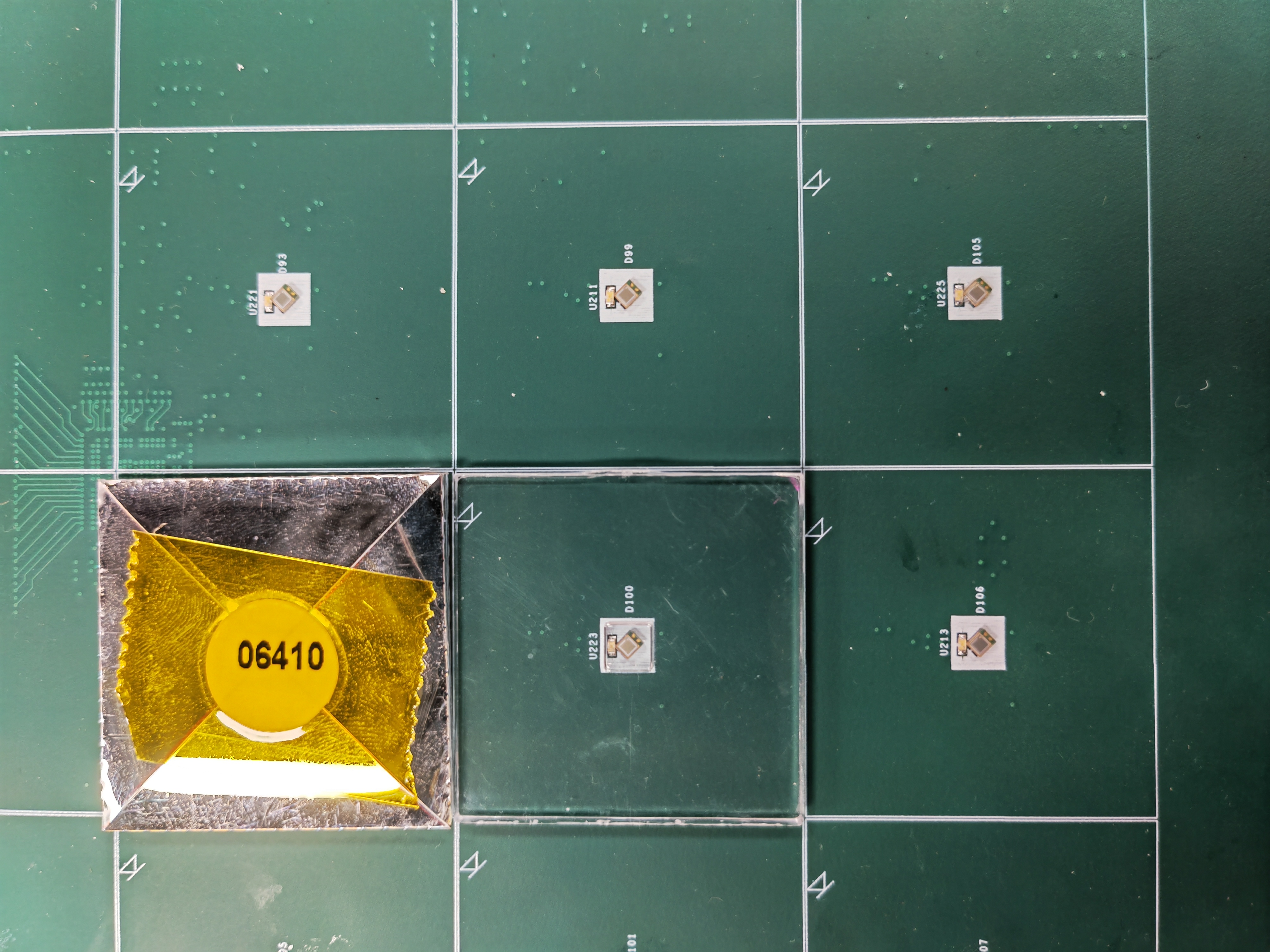}
          \label{Fig:HBU1}
    }
    \subfigure[]
    {
            \centering
            \includegraphics[width=.35\textwidth]{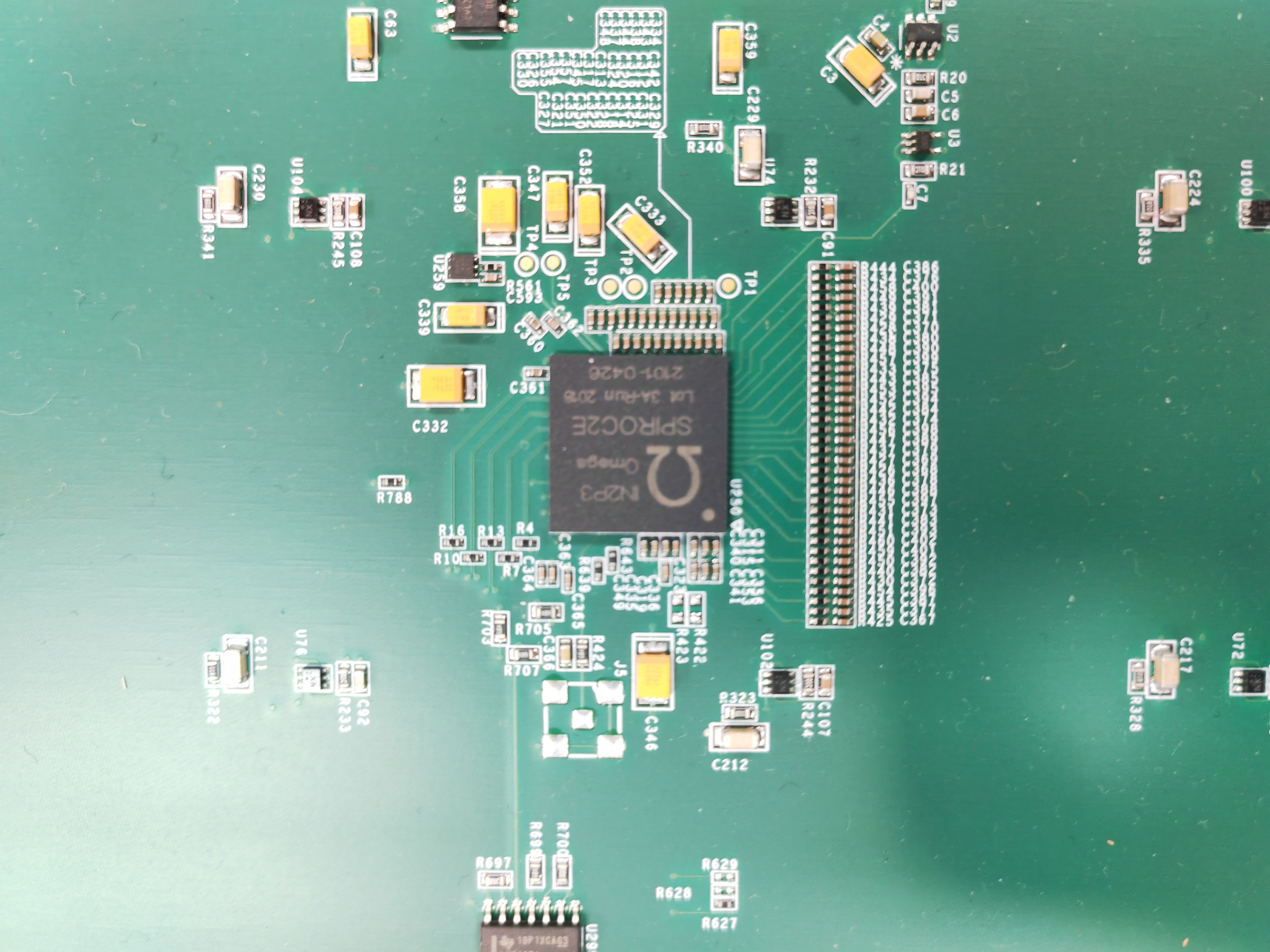}
          \label{Fig:HBU2}
    }
    \caption{(a) Overview of the electronic PCBs: the DAQ board(top left), the HBU without scintillator(top right), the HBU assembled with scintillator(bottom left), the electronic side of the HBU and the DIF board(bottom right). (b) The sensitive unit on HBU. (c) The SPIROC2E chip on HBU.}
\end{figure}

    The readout electronic system is responsible for collecting data from 12,960 channels across the 40 sensitive layers. It is composed of the front-end electronics and the DAQ system. The front-end electronics consists of 120 HBU boards and the DAQ system consists of 40 Data InterFace(DIF) boards and a DAQ board, as shown in \cref{Fig:Boards}. The HBU board is responsible for carrying the sensitive units and converting the analog signals of the sensitive units into digital signals, the DIF board is designed to collect the digital signals across the sensitive layer, and the DAQ board is intended to gather the data from the DIF boards and transmit it to the server. 

    On one side of the HBU, 108 SiPMs were soldered with LEDs positioned beside them, and the scintillator tiles were subsequently affixed using glue, as depicted in \cref{Fig:HBU1}. On the opposite side of the HBU, three SPIROC2E chips were employed for the data collection of these 108 sensitive units, as illustrated in \cref{Fig:HBU2}.

\begin{figure}[htbp]
        \centering
        \subfigure[]
        {
                \centering
                \includegraphics[width=.35\textwidth]{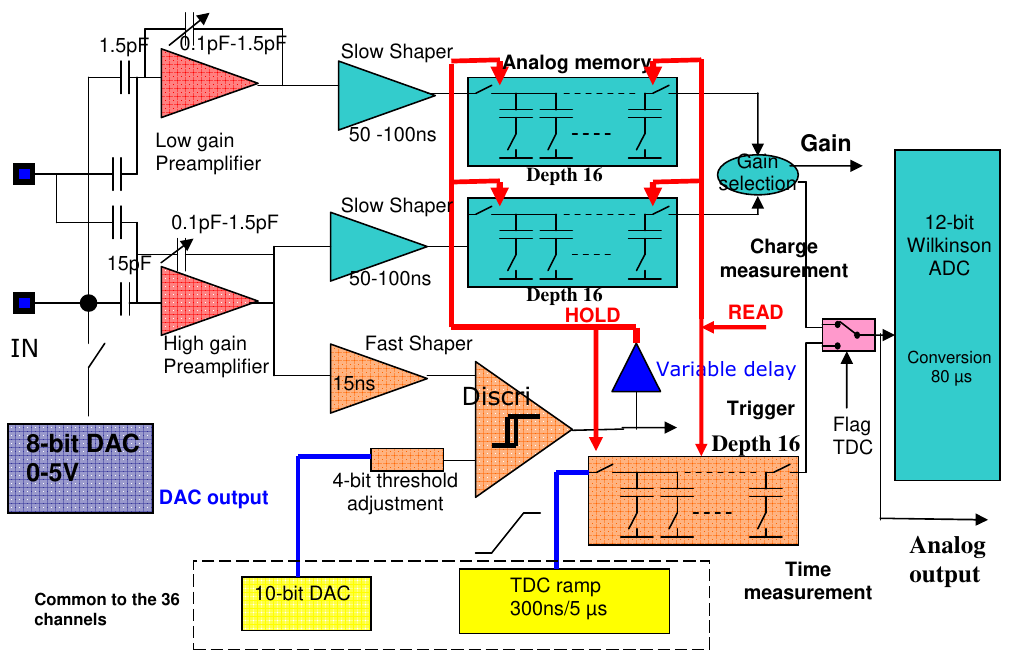}
                \label{Fig:SPIROC}
        }
        \subfigure[]{        
            \includegraphics[width=.35\textwidth]{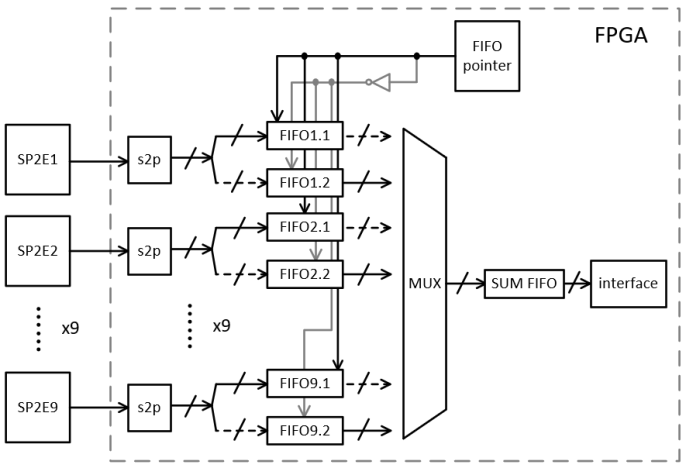}
            \label{Fig:HBU_parallel}
        }
        \subfigure[]{        
            \includegraphics[width=.35\textwidth]{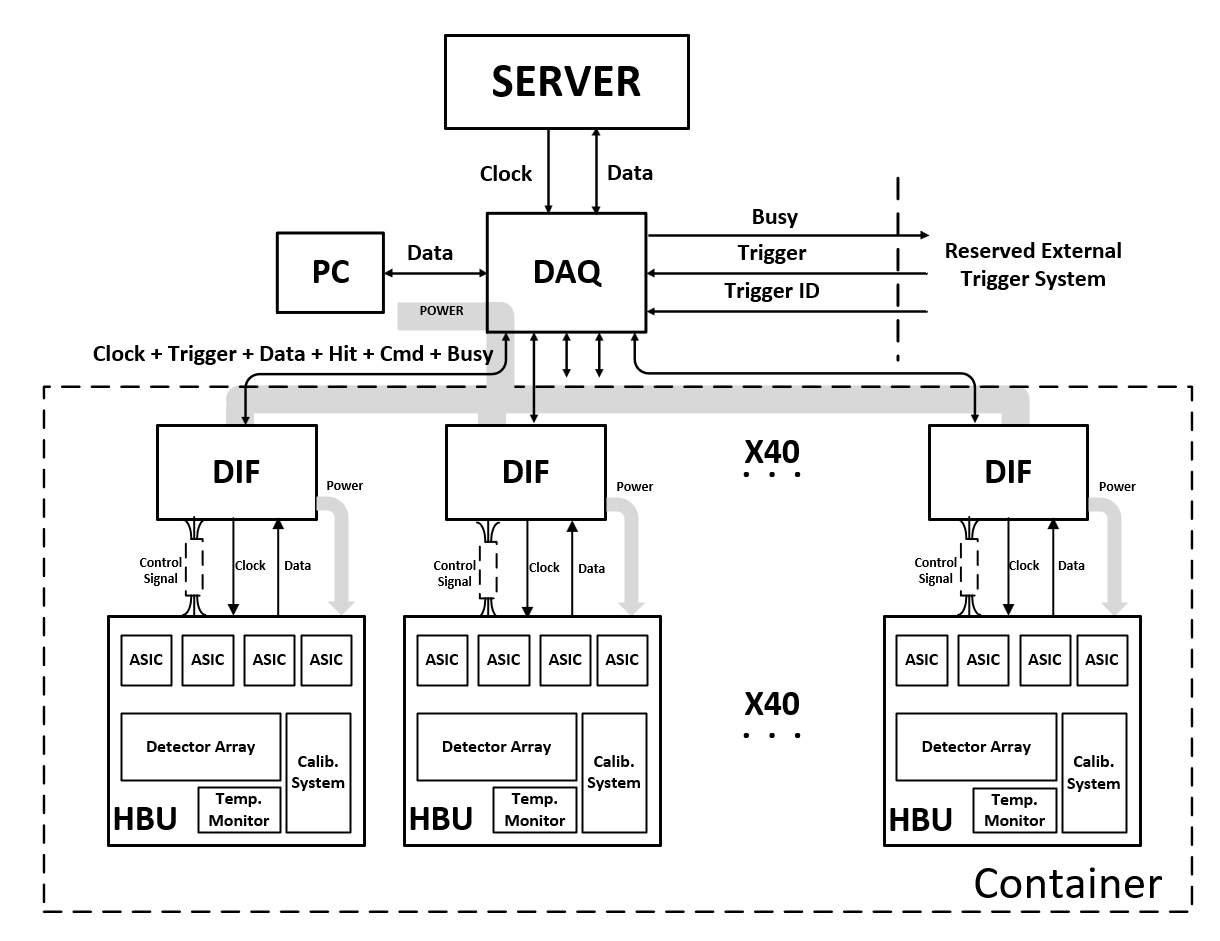}
            \label{Fig:HBU3}
        } 
        \caption{(a)The design of the SPIROC2E chip, including 36 analog channels and a 12-bit Wilkinson ADC\cite{SPIROC2}. (b) Parallel readout scheme of 9 SPIROC2E chips in a single AHCAL layer\cite{Zhou_2023}. (c) Schematic view of the readout and data acquisition, involving 40 DIF boards and a DAQ board.}
    \end{figure}

    SPIROC is an auto-triggered, bi-gain, 36-channel ASIC which allows to measure on each channel the charge from one photoelectron to 2000 and the time with a \SI{100}{ps} accurate TDC\cite{SPIROC1,ConfortiDiLorenzo:2013vka,SPIROC2}. An analog memory array with a depth of 16 for each is used to store the time information and the charge measurement. A 12-bit Wilkinson ADC has been embedded to digitize the analog memory contents (time and charge on 2 gains). The data are then stored in a \SI{4}{kbytes} RAM. A very complex digital part has been integrated to manage all theses features and to transfer the data to the DAQ. Each SPIROC chip contains 36 channels, corresponding to 36 sensitive units. Each SPIROC channel employs two preamplifiers with different gains to enhance the dynamic range, as depicted in \cref{Fig:SPIROC}. Following the high gain preamplifier, a fast shaper and a discriminator are used to provide self-trigger. Once triggered, the signals from the high gain preamplifier and the low gain preamplifier are recorded in the analog memory after the slow shaping. Subsequently, the signals stored in the analog memory are converted into digital signals by a 12-bit ADC. It is important to note that the signal that triggers the SPIROC chip may not be exactly the same as the signal recorded. For instance, if a signal with an extremely narrow time width triggers the discriminator after the fast shaper, a pedestal may be recorded after the slow shaper.

    A single sensitive layer of AHCAL contains 3 HBUs, corresponding to 324 sensitive units. A DIF board was designed to collect signals from these HBUs with a parallel readout scheme, as illustrated in \cref{Fig:HBU_parallel}. 18 small FIFO (First In/First Out) modules and one large FIFO module were designed to cache the data from 9 SPIROC2E chips, enabling the simultaneous processing of the Analog to Digital (AD) conversion and the data transmission. This parallel readout scheme significantly enhanced the event readout rate to \SI{3}{kHz}, comparing to the \SI{83}{Hz} of serial readout scheme, enabling the prototype to operate under the beamline environment\cite{Zhou_2023}. The FPGA on the DIF board also processes commands from upstream, real-time control of the HBU board, and power supply to the HBU boards.

    A DAQ system was designed to collect the data from 40 DIF boards and transmit the data to the server or the PC. Additionally, it synchronizes the clocks across the 40 AHCAL layers, and dispatches triggers and other commands to the DIF boards. \cref{Fig:HBU3} provides a schematic view illustrating the interactions between the front-end electronics and the DAQ system.

    The trigger system of the AHCAL consists of the SPIROC2E chips, the DAQ board and the trigger logic unit (TLU)\cite{TLU}. The TLU distributes the trigger based on the coincidence measurement, which is determined by signals from other detectors or two specific layers of the AHCAL prototype. Once receives the trigger from the TLU, the DAQ sends a readout command to all SPIROC2E chips via DIF boards. The SPIROC2E chips start the AD conversion and the FIFO cache starts to readout. If the absence of the trigger is over \SI{4}{\upmu s}, the DAQ board sends an erase command to clear the cache.

\subsection{Calibration system}
\label{Calibration_system}
    
    The non-uniformities among the numerous AHCAL channels would significantly impact the performance of the AHCAL. Therefore, a calibration system was developed. The SPIROC chips were modified to calibrate the pedestal and gain of each channel. SiPMs were coupled with LEDs for gain monitoring. Additionally, each sensitive layer was equipped with 48 temperature sensors to monitor temperature variations.

\begin{figure}[htbp]
        \centering
        \subfigure[]{
            \includegraphics[width=.35\textwidth]{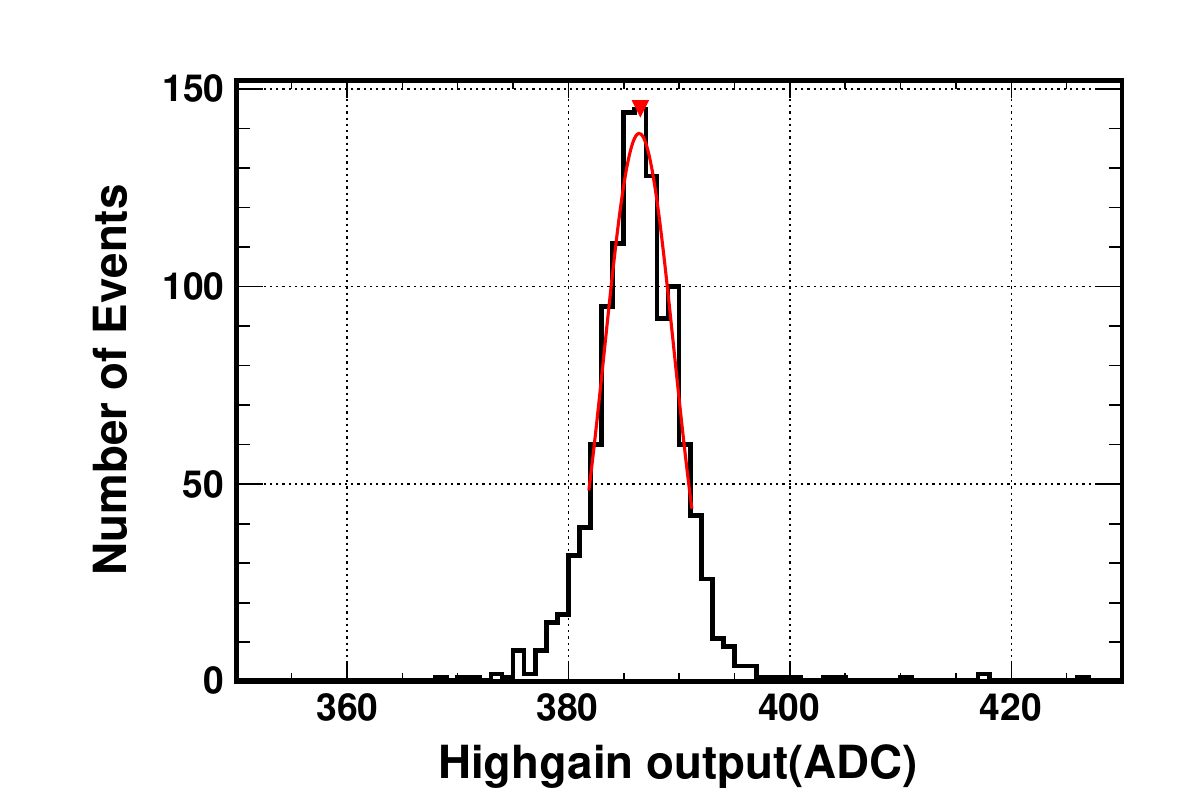}
            \label{Fig:pedestal_fit}
        }
        \subfigure[]{
            \includegraphics[width=.35\textwidth]{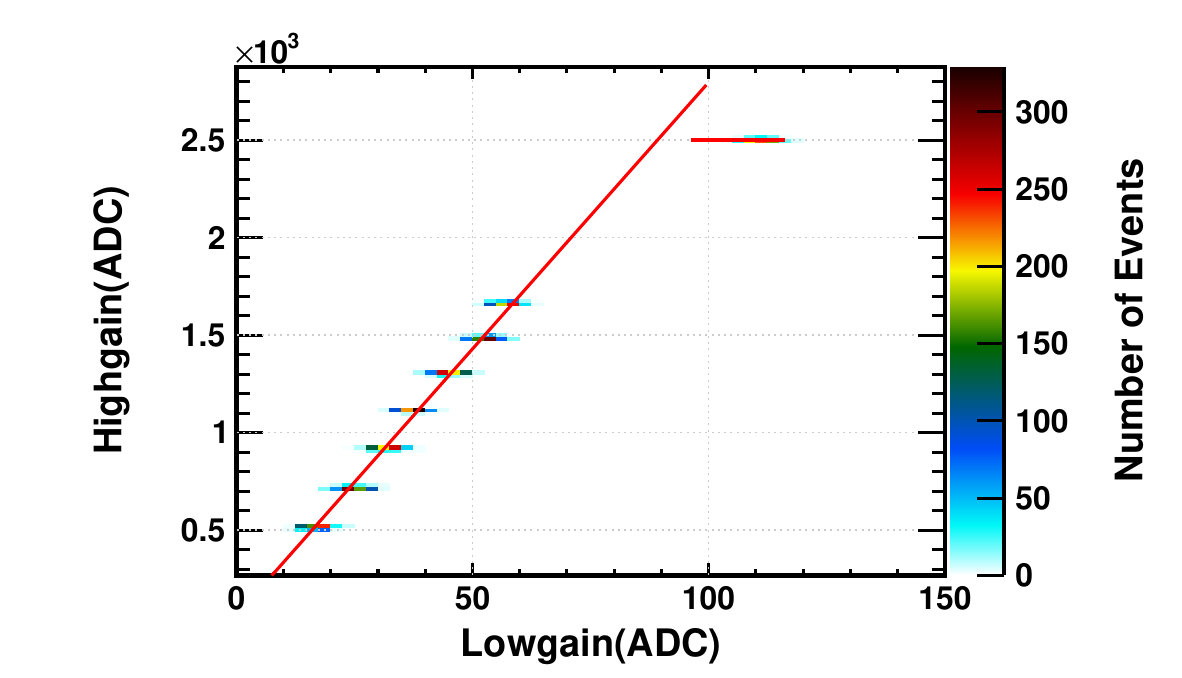}
            \label{Fig:GR_fit}  
        }
        \subfigure[]{
            \includegraphics[width=.35\textwidth]{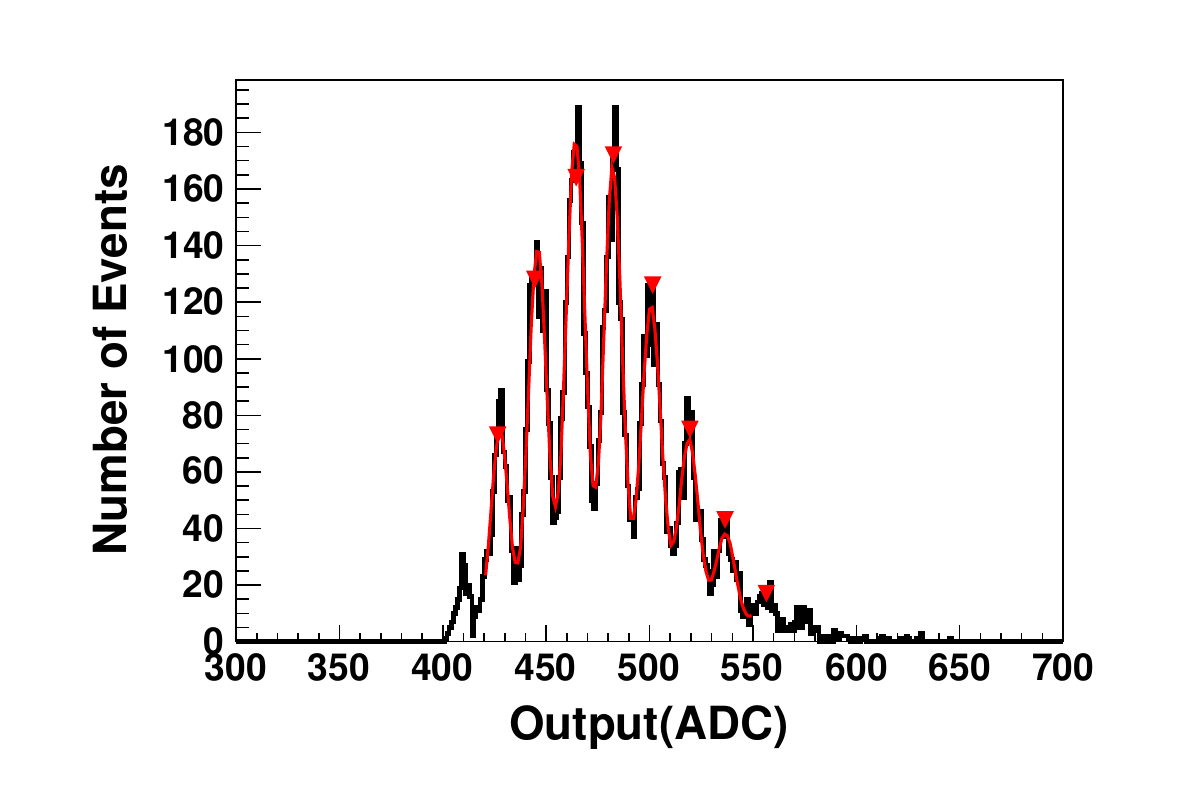} 
            \label{Fig:LED} 
        }
        \caption{(a) Distribution of pedestal for a single AHCAL channel, fitted by a Gaussian. (b) Low and high gain calibration for a single AHCAL channel, fitted by 2 linear function. (c) LED spectrum of a single AHCAL channel, each peak corresponds to a photoelectron.}
    \end{figure}

    The pedestal of each channel could be obtained by a random trigger distributed from the DIF board to SPIROC chips. The pedestal of this channel is determined as the mean value derived from the Gaussian fitting, as illustrated in \cref{Fig:pedestal_fit}.

    The DIF board could also inject electric charge into the SPIROC chip to probe the response of all channels. By varying the amount of injection charge, the gain ratio between high gain and low gain could be calibrated, as show in \cref{Fig:GR_fit}. The high gain response is linear with the low gain response till it reaches its saturation point. With the linear part and the saturation part fitted separately, gain ratio could be obtained from the slope of the left line, and the saturation point is considered to be the intersection of two lines. 

    The LED placed adjacent to the SiPM is used to calibrate and monitor the SiPM. On a single HBU, 108 LEDs are divided into two groups to prevent light crosstalk. The SiPM response to the LED exhibits good single photon separation as illustrated in \cref{Fig:LED}. The intervals between photon peaks are approximately \SI{20}{ADC}, representing the gain of the SiPM. 
   
    The performances of SiPMs such as gain, PDE, and cross talk were notably influenced by temperature. Therefore, a temperature monitoring system consisting of 48 temperature sensors on each sensitive layer was implemented. 
	
\subsection{Mechanical structure}
    
    The mechanical structure of the AHCAL prototype is composed of the steel cassettes and the supporting framework. The steel cassette was designed to accommodate three HBUs and one DIF board. The supporting framework is responsible for securing the steel cassettes and absorber plates.

\begin{figure}[htbp]
        \centering
        \subfigure[]{
            \includegraphics[width=.35\textwidth]{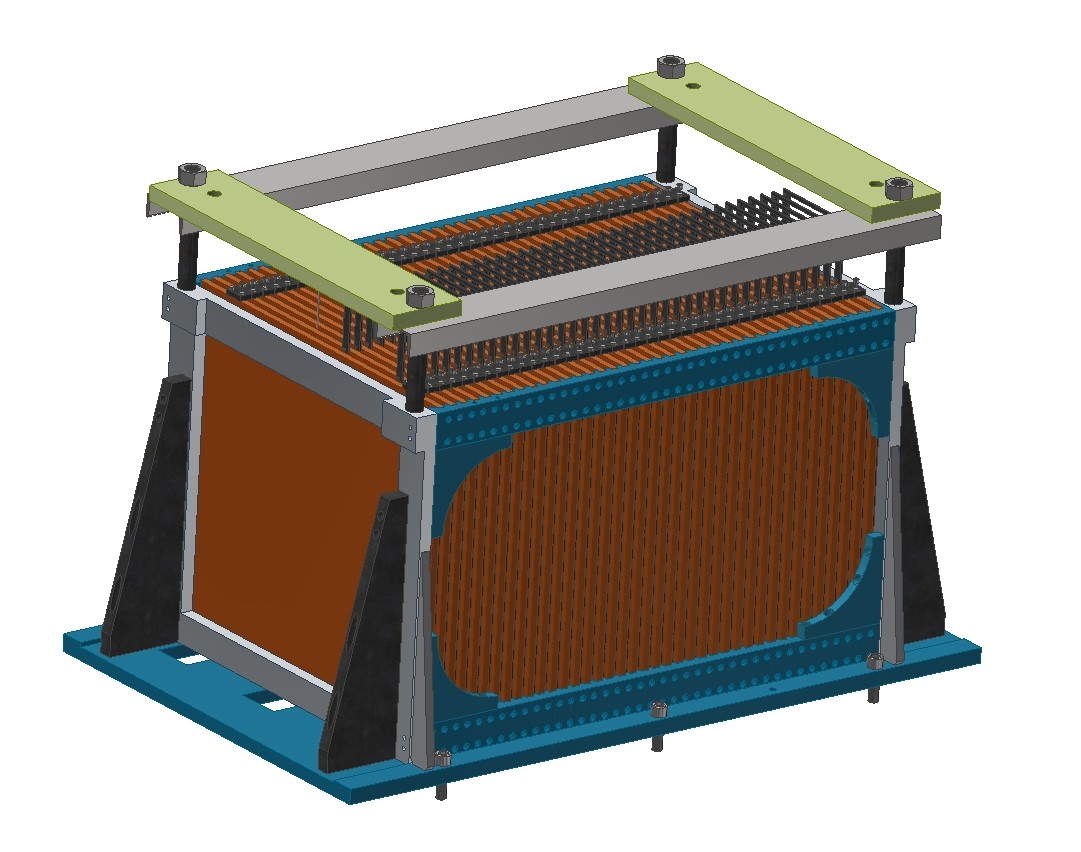}
            \label{Fig:mechanics_prototype}
        }
        \subfigure[]{
            \includegraphics[width=.35\textwidth]{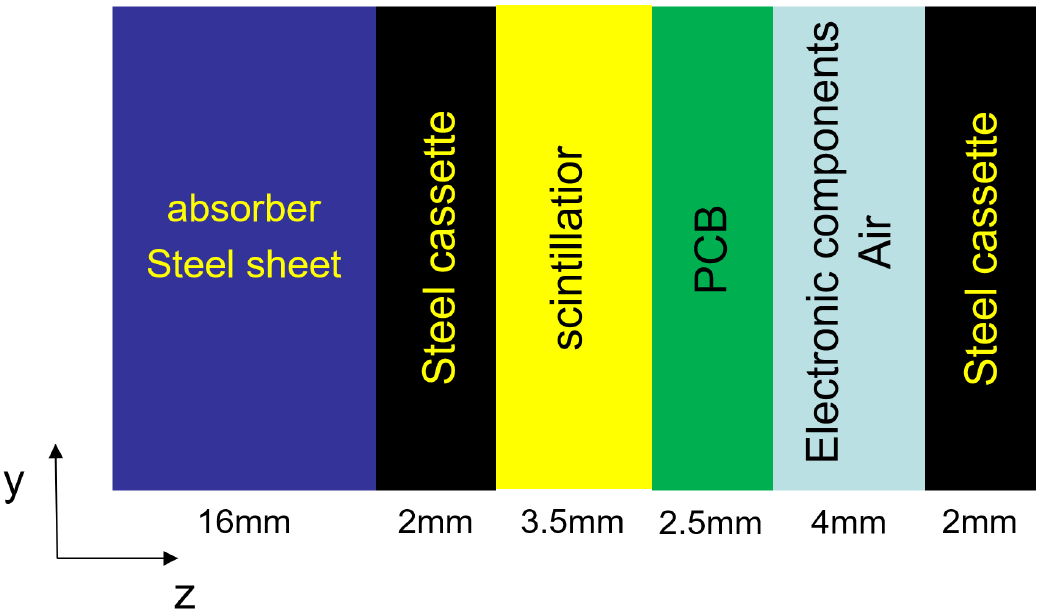}
            \label{Fig:cross_section} 
        }
        \subfigure[]
        {
            \includegraphics[width=.35\textwidth]{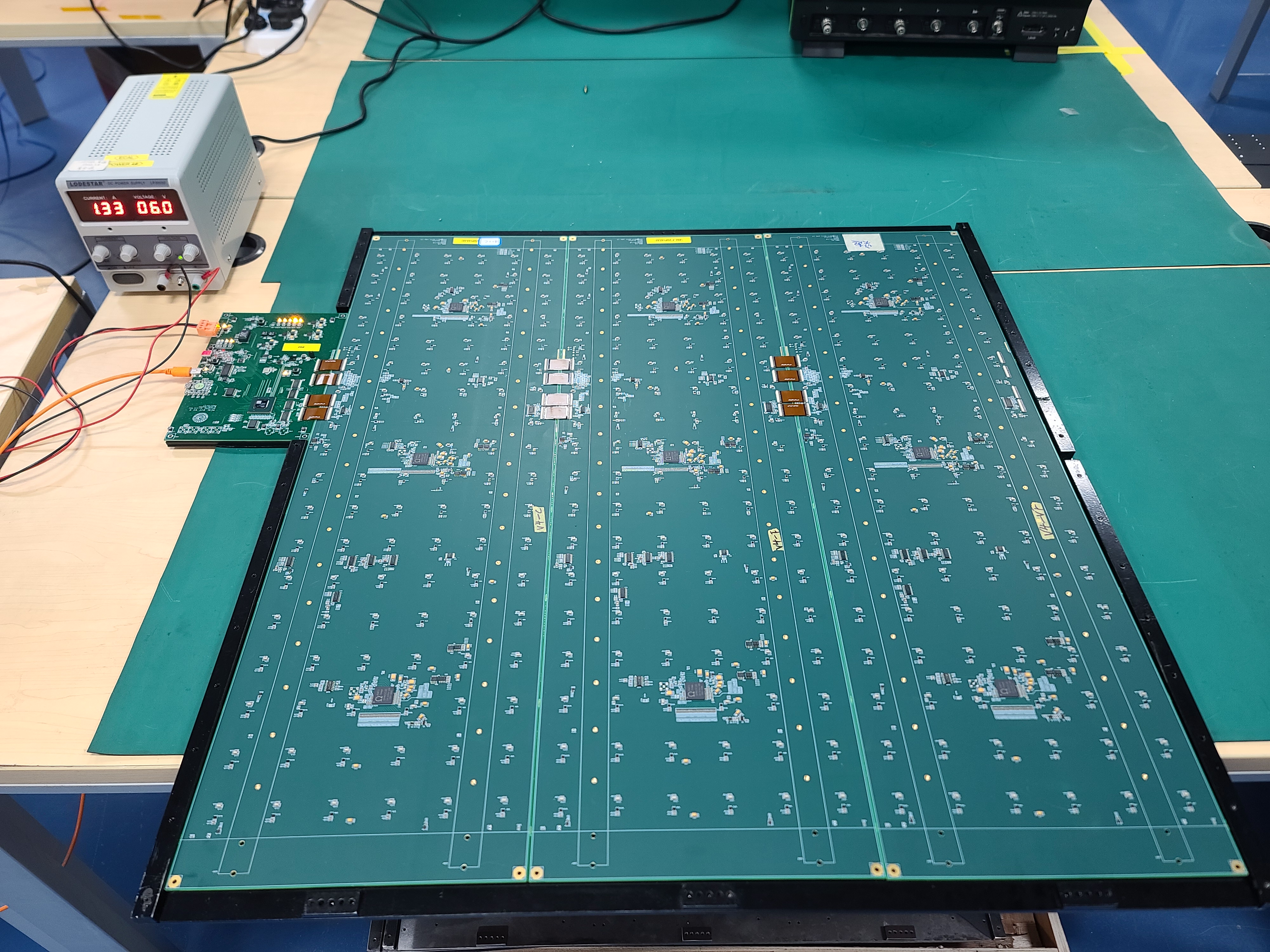}
            \label{Fig:mechanics_cassette}
        }
        \caption{(a) Mechanical scheme of the prototype structure, 40 sensitive layers and absorbers were fixated.(b) A schematic cross-section of the AHCAL prototype.(c) The steel cassette assembled with three HBUs and the DIF board.}
        
    \end{figure}

    The design of the AHCAL prototype supporting framework is depicted in \cref{Fig:mechanics_prototype}. This framework supports and secures 40 absorber plates measuring \SI{16}{mm}, with steel cassettes accommodating the scintillator tiles and electronics inserted into the gaps between these plates. The total thickness of the absorbing material is \SI{800}{mm}, which includes contributions from the steel cassettes.

	The design of the steel cassettes pursuits of the compactness of the AHCAL prototype, As illustrated in \cref{Fig:cross_section}. The top and bottom steel sheets of the cassettes, which serve as part of the absorbing material, are both \SI{2}{mm} thick. This thickness ensures the stiffness of the cassette while maintaining portability. The scintillator tile, wrapped with ESR, has a thickness of \SI{3.5}{mm}, while the PCB has a thickness of \SI{2.5}{mm}. Additionally, a \SI{4}{mm} space is designed for the electronic parts, providing a tolerance of \SI{1}{mm}. The total thickness of the cassette is \SI{14}{mm}. 

    \cref{Fig:mechanics_cassette} illustrates a steel cassette assembled with three HBUs and the DIF board. The scintillator tiles are in direct touch with the bottom sheet, and six steel strips with a thickness of \SI{3.5}{mm} are fixed to the bottom sheet secure the scintillator tiles and provide screw holes for HBU fixation. This design provides light shielding to the scintillator tiles while maintaining the compactness as much as possible. The HBUs will be attached to the top sheet with six small strips, leaving space for air cooling.  

\section{Construction of the AHCAL prototype}

    In the construction process of the AHCAL prototype, a total of 16000 scintillator tiles were produced and tested, along with 120 HBU boards that were manufactured and soldered with SiPMs. They were assembled into 40 sensitive layers with other components. Furthermore, 40 DIF boards, the DAQ board, 40 absorber plates, and the supporting framework were fabricated. After the integration of all these components, the AHCAL prototype was completed.

\subsection{Production of sensitive units}
\begin{figure}[htbp]
        \centering
        \subfigure[]{
            \includegraphics[width=.35\textwidth]{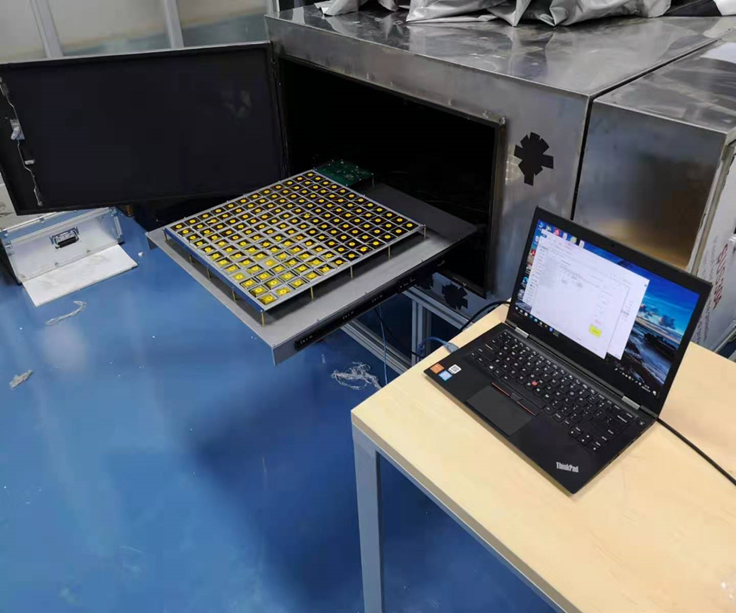}
            \label{Fig:ScB3}
        }
        \subfigure[]{
            \includegraphics[width=.35\textwidth]{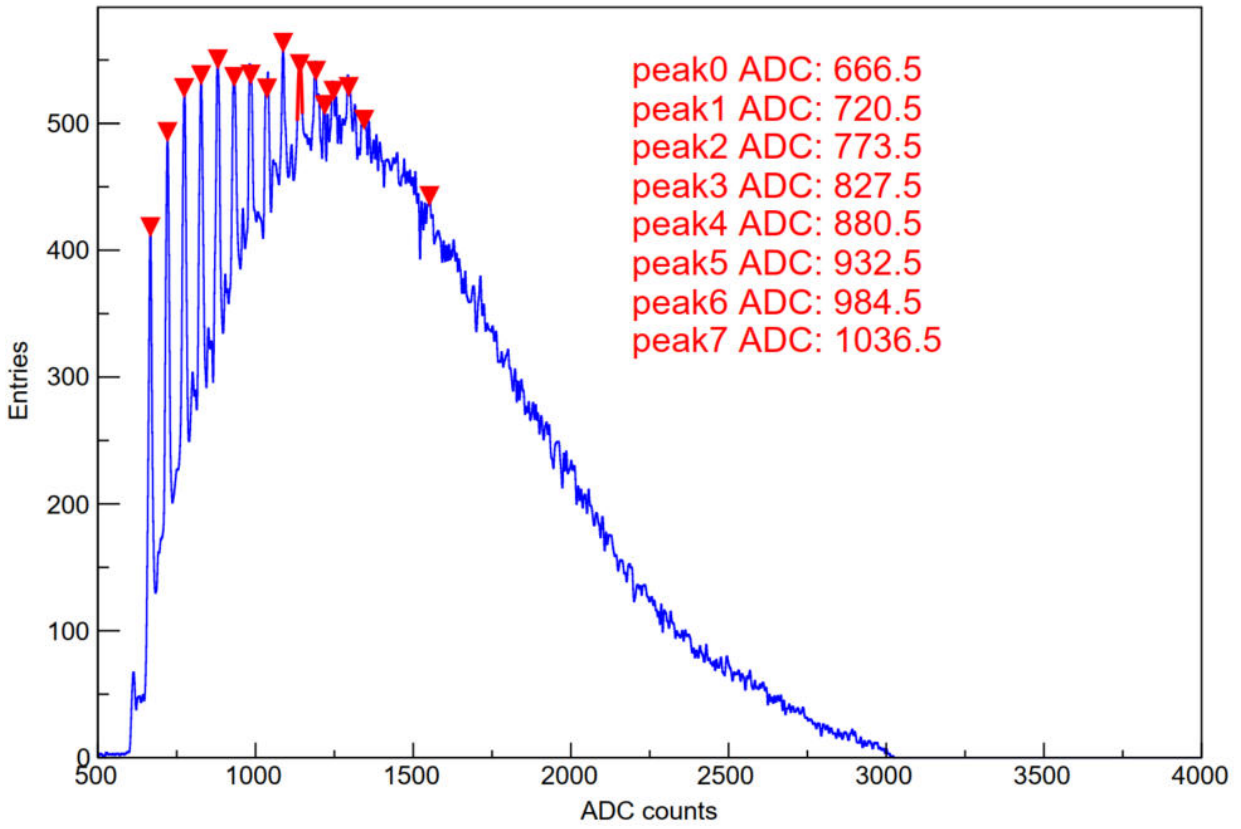}
            \label{Fig:ScB4}
        }
        \subfigure[]{
            \includegraphics[width=.35\textwidth]{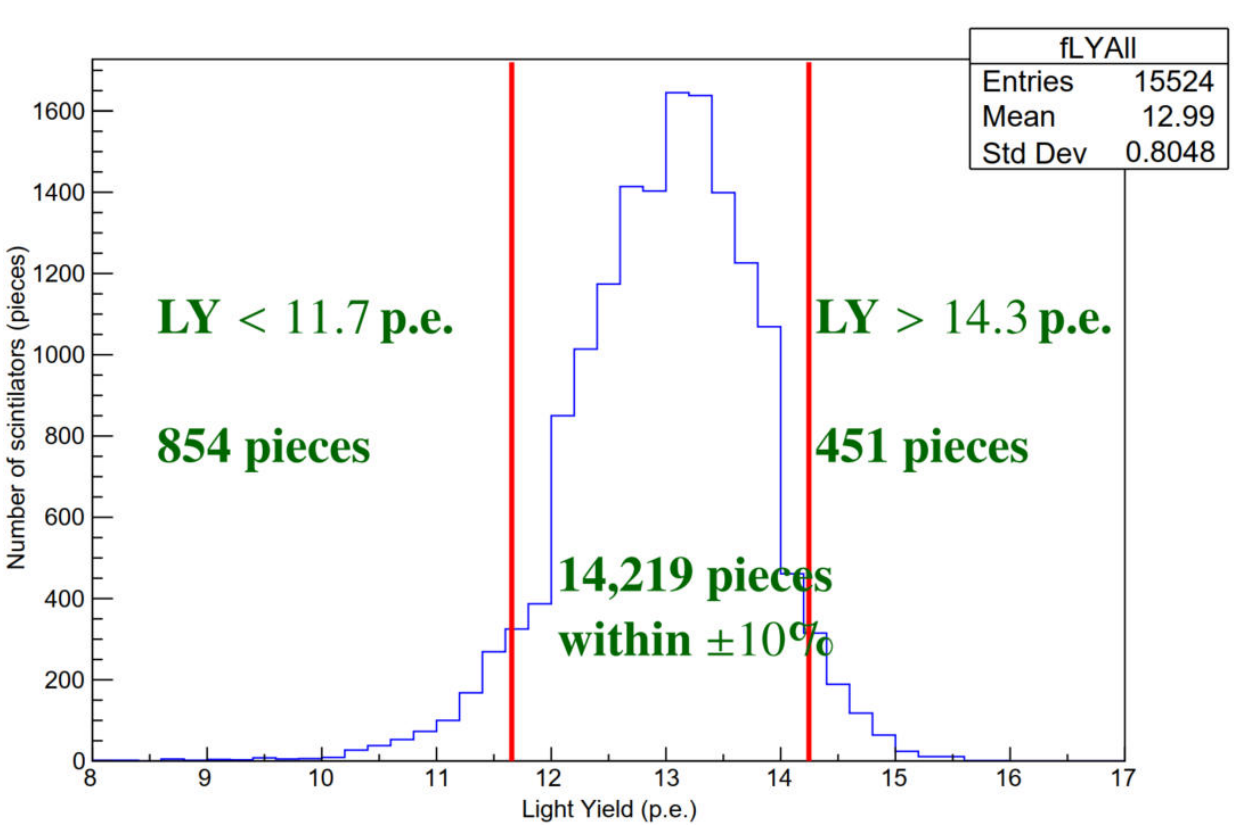}
            \label{Fig:ScB6}
        }
        \caption{(a) The test platform for scintillator tiles. (b) The MIP spectrum of a single scintillator tile, each peak corresponds to a photoelectron\cite{duan2022scintillator}. (c) Light yield distribution of all 15524 scintillator tiles}
    \end{figure}

	The scintillator tiles were produced efficiently with the injection molding technique and wrapped with ESR foils using a dedicated machine.
    A light yield test platform was developed to test this vast quantity of scintillator tiles\cite{liu2020development}. The platform consists of a front-end board, a DIF board, a Sr-90 source loaded on a 3D servo motor, and a PC, as demonstrated in \cref{Fig:ScB3}. The front-end board followed the design of HBU, utilizing 144 S13360-1325PE SiPMs and 4 SPIROC2E chips, allowing for the testing of 144 scintillator tiles in a single run.

    \cref{Fig:ScB4} displays a MIP spectrum for a single scintillator tile. The peaks corresponding to photon electrons were identified, and the ADC value of a single photon-electron was estimated. The light yield of the scintillator tile was determined by the most probable value (MPV) obtained from fitting a convolution of the Landau and Gaussian functions.
 
    \cref{Fig:ScB6} displays the distribution of light yield values for all the scintillator tiles. To improve the uniformity of the AHCAL prototype, only scintillator tiles with a light yield falling within a window of $13\pm10\%$ p.e. were selected. Out of approximately 16000 scintillator tiles, 14219 were chosen under this criterion\cite{duan2022scintillator}.
   
    A SiPM test platform was developed based on customized jigs and LEDs. The customized jig can contain four SiPMs with a hole in the middle, allowing the LED light to pass through and activate the SiPMs. 
    An FPGA was utilized to control the LED and readout signals from the SiPMs. Good photon electron separation was achieved on this platform, similar as what has been achieved with the LED calibration, as shown in \cref{Fig:LED}. SiPMs produced in the same batch share similarities, such as the working voltage, while SiPMs from different batches differ in these characteristics. Therefore, in every batch, a few of SiPMs were tested with the SiPM test platform to ensure their quality.


\subsection{Assembly of sensitive layers}


    %
    %

        The HBU boards were produced and soldered with SiPMs, SPIROC chips and other components. The scintillator tiles were assembled and secured onto the PCB using adhesive. The HBUs along with scintillator tiles were secured in the steel cassette, as illustrated in \cref{Fig:mechanics_cassette}. Prior to installing the top steel sheet, an electronic test was conducted to verify the basic functionality of this sensitive layer. The pedestals of all channels were checked. LED was activated group by group as to find out potential light crosstalk or dead channels. 


\begin{figure}[htbp]
    \centering
    \includegraphics[width=.35\textwidth]{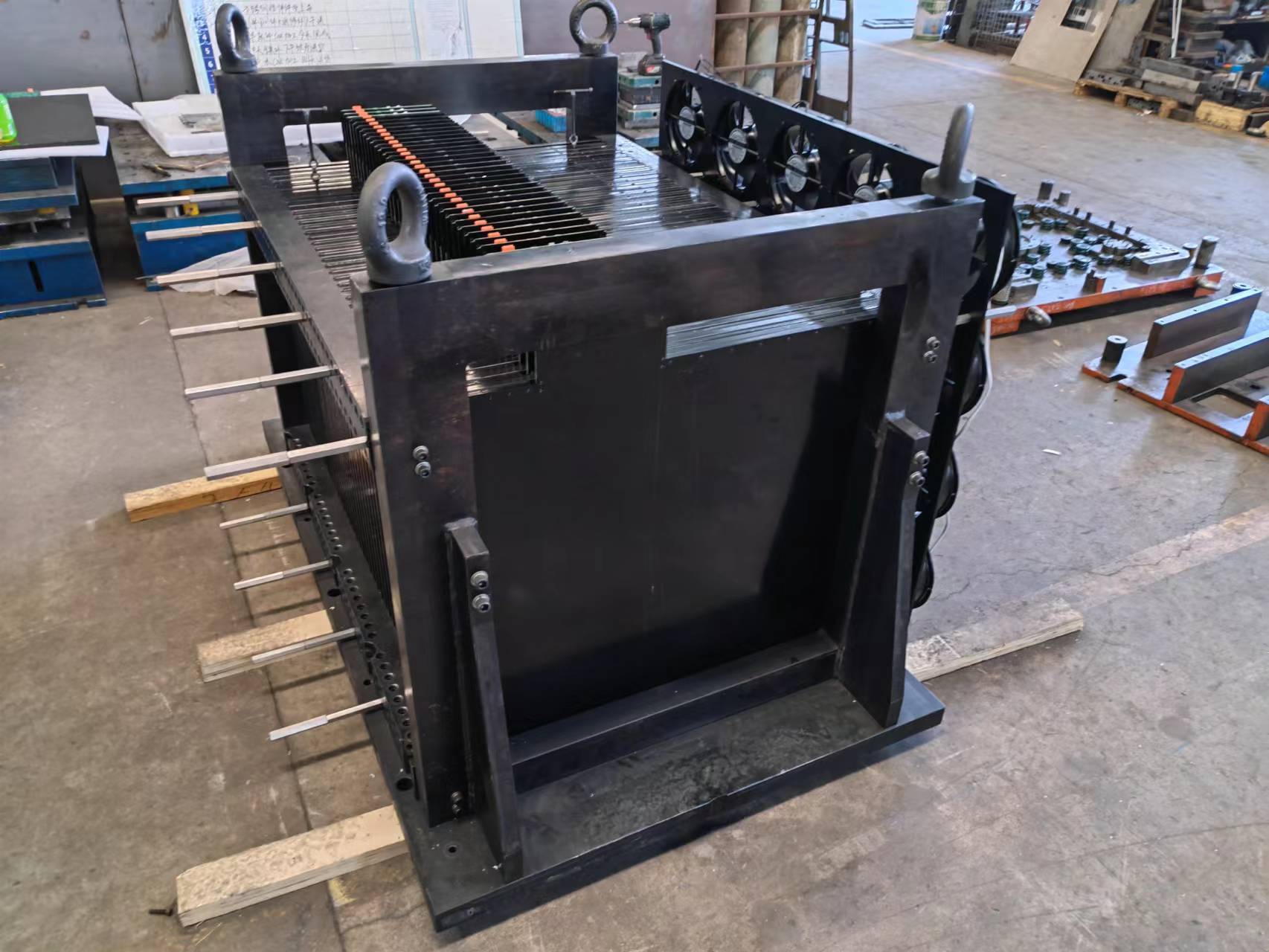}
    \caption{The integrated AHCAL prototype.}
    \label{Fig:mechanics_prototype4}
\end{figure}	

\section{Tests of the AHCAL prototype}

\subsection{Integration of the prototype}

    The supporting framework of the prototype was manufactured and then integrated with the steel absorbing plates, the gaps between the absorbers were tested to ensure the tolerance for sensitive layers. The sensitive layers were inserted layer by layer. Several fans were fixed on the right side of the prototype for air cooling, as depicted in \cref{Fig:mechanics_prototype4}. The complete AHCAL prototype reaches a total weight of 5.5 tons.

 In order to calibrate the pedestal and gain ratio of each AHCAL channel, an electronic test was conducted. Additionally, a cosmic ray test was performed to validate the functionality of the entire prototype.

\subsection{Electronic test}
    
\begin{figure}[htbp]
        \centering
        \subfigure[]{
            \centering
            \includegraphics[width=.35\textwidth]{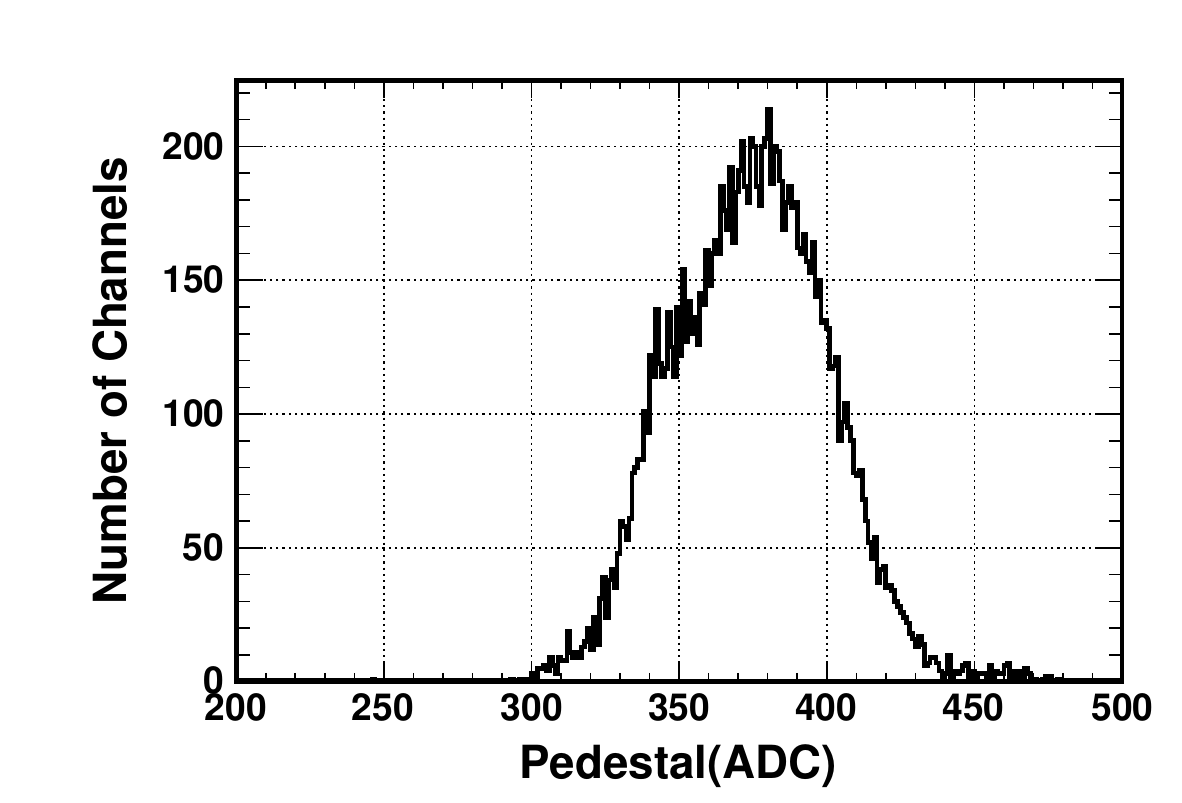}
        }
        \subfigure[]{
            \centering
            \includegraphics[width=.35\textwidth]{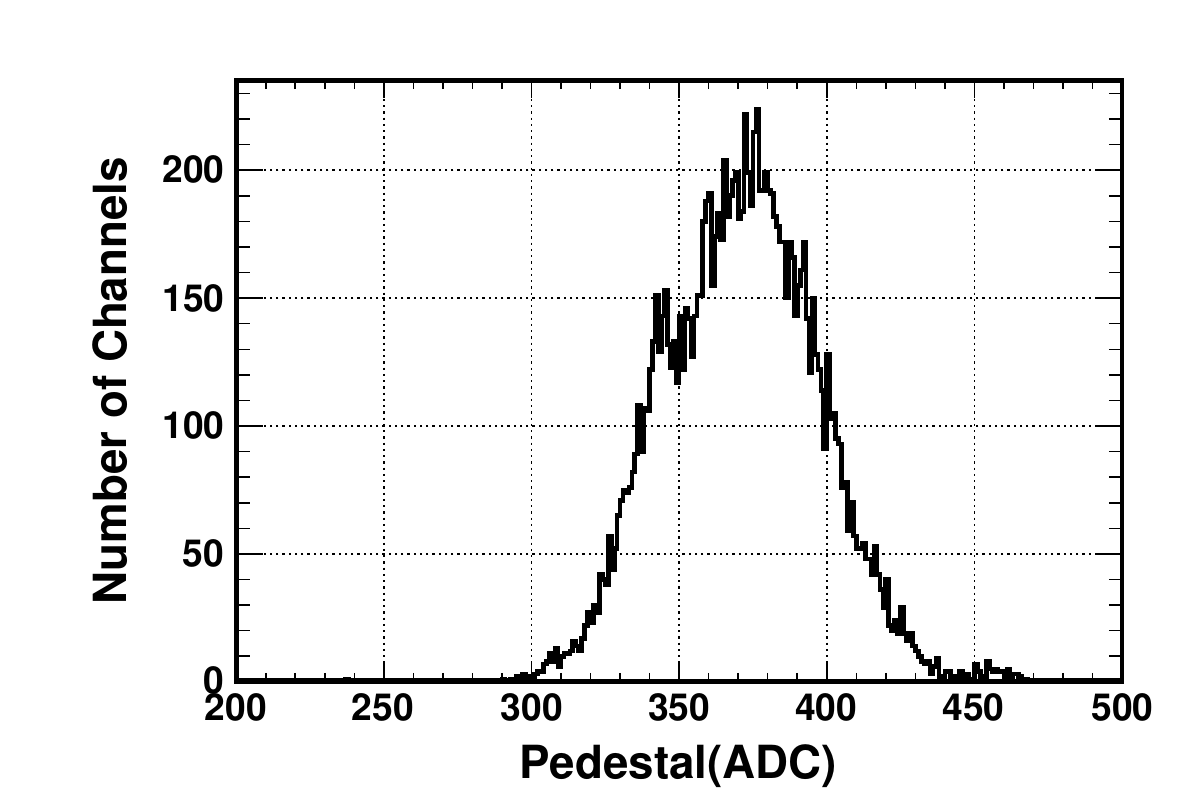}
        }
        \subfigure[]{
            \centering
            \includegraphics[width=.35\textwidth]{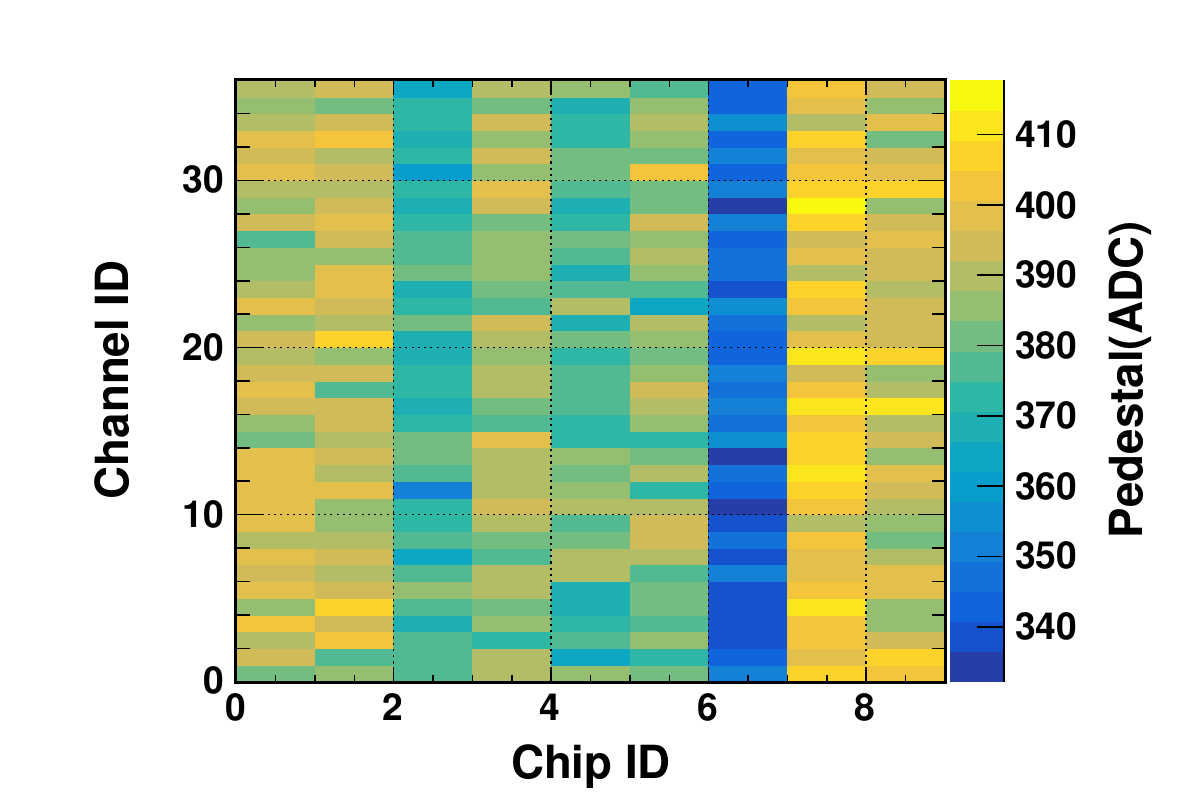}
            \label{Fig:pedestal_map}
        }
        \caption{Pedestals for AHCAL channels: (a) high gain channels (b) low gain channels. (c) Pedestals from different chips in a sensitive layer, with obvious differences on pedestals between chips.}
        \label{Fig:pedestal}
    \end{figure}

    The electronic test was carried out using the calibration system introduced in Section \ref{Calibration_system}. The pedestals of both high gain channels and low gain channels were obtained through an external trigger. \cref{Fig:pedestal} depicted the distributions of pedestals for all AHCAL high gain channels and low gain channels.

    \cref{Fig:pedestal_map} depicts the pedestals of different chips in a sensitive layer. Consistency within a chip and differences among chips were observed, indicating that non-uniformities among pedestals mainly originated from the variations among SPIROC chips.  

\begin{figure}[htbp]
        \centering
        \subfigure[]{
            \centering
            \includegraphics[width=.2\textwidth]{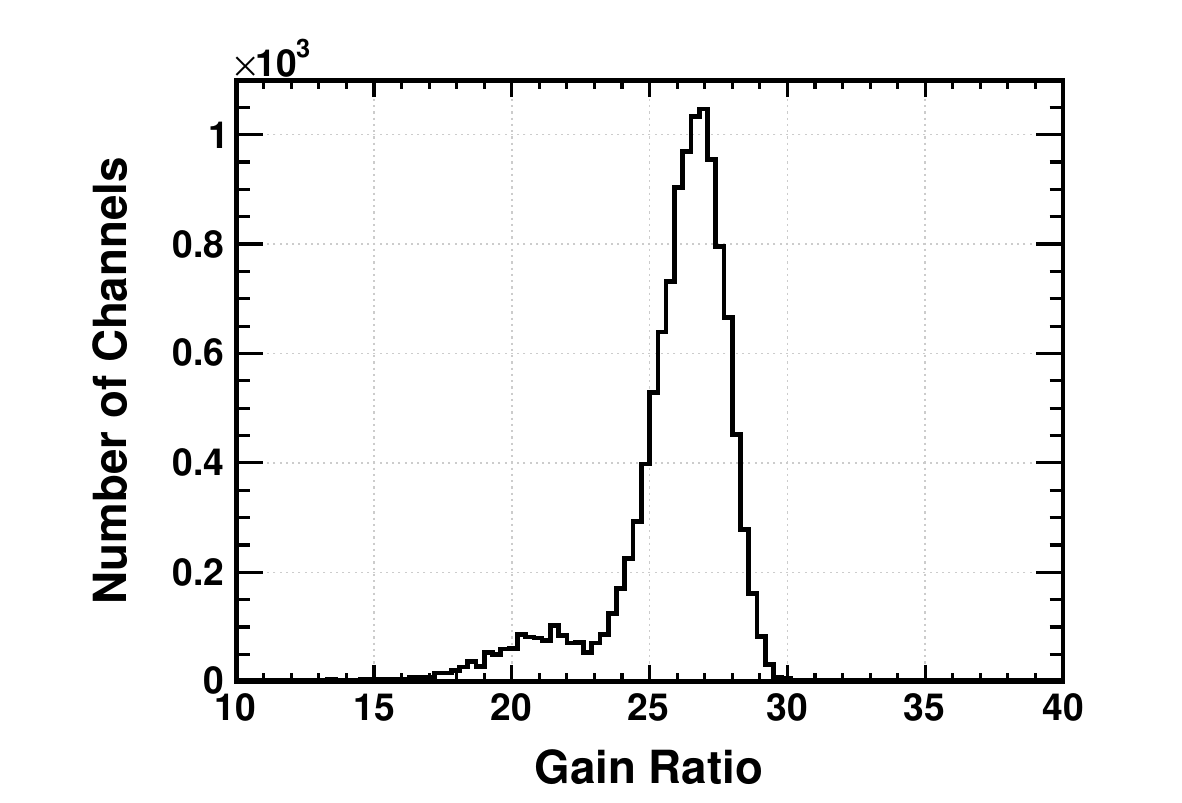}
        }
        \subfigure[]{
            \centering
            \includegraphics[width=.2\textwidth]{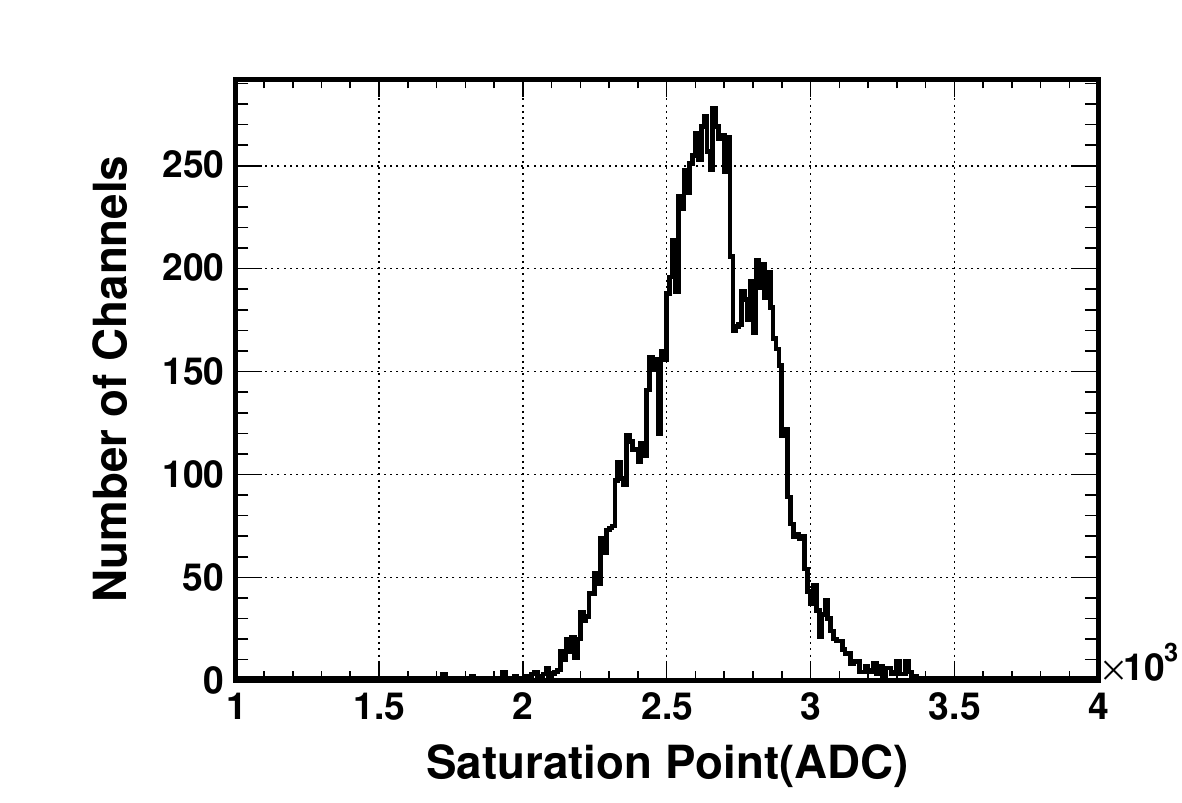}
        }
        \subfigure[]{
            \centering
            \includegraphics[width=.2\textwidth]{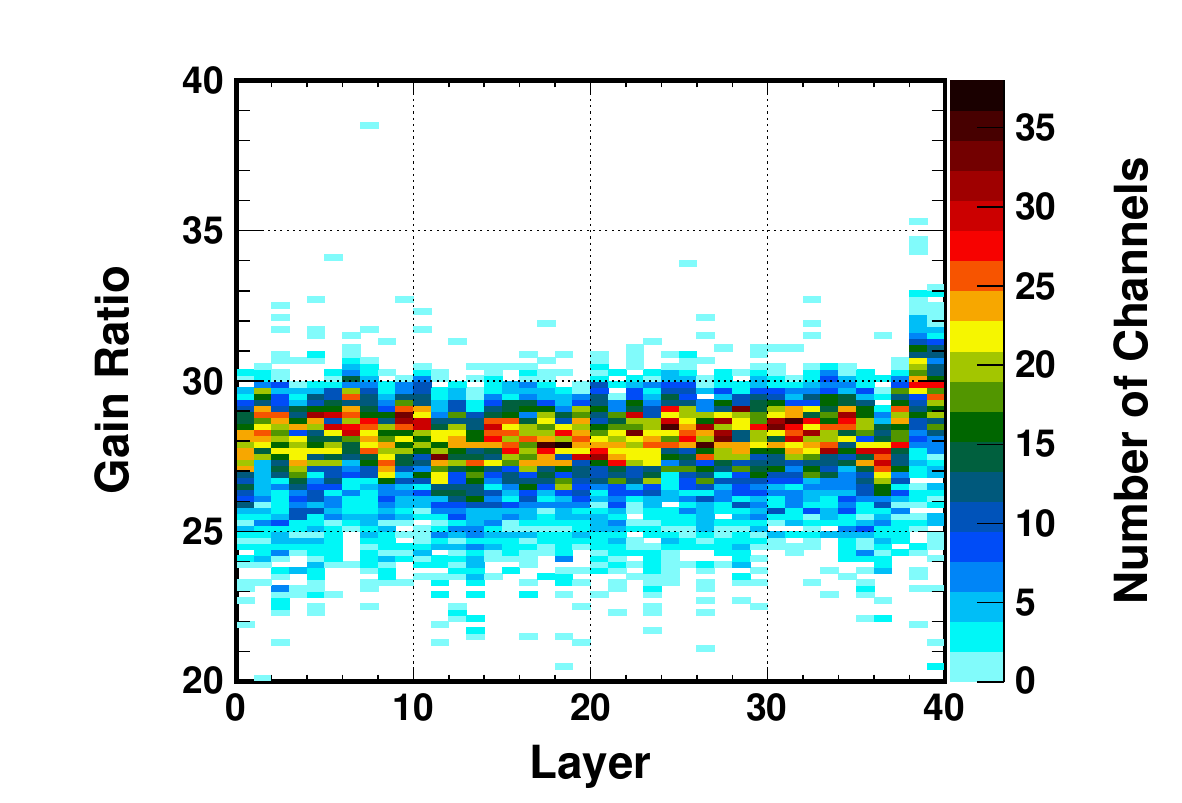}
            \label{Fig:GR_layer}
        }
        \subfigure[]{
            \centering
            \includegraphics[width=.2\textwidth]{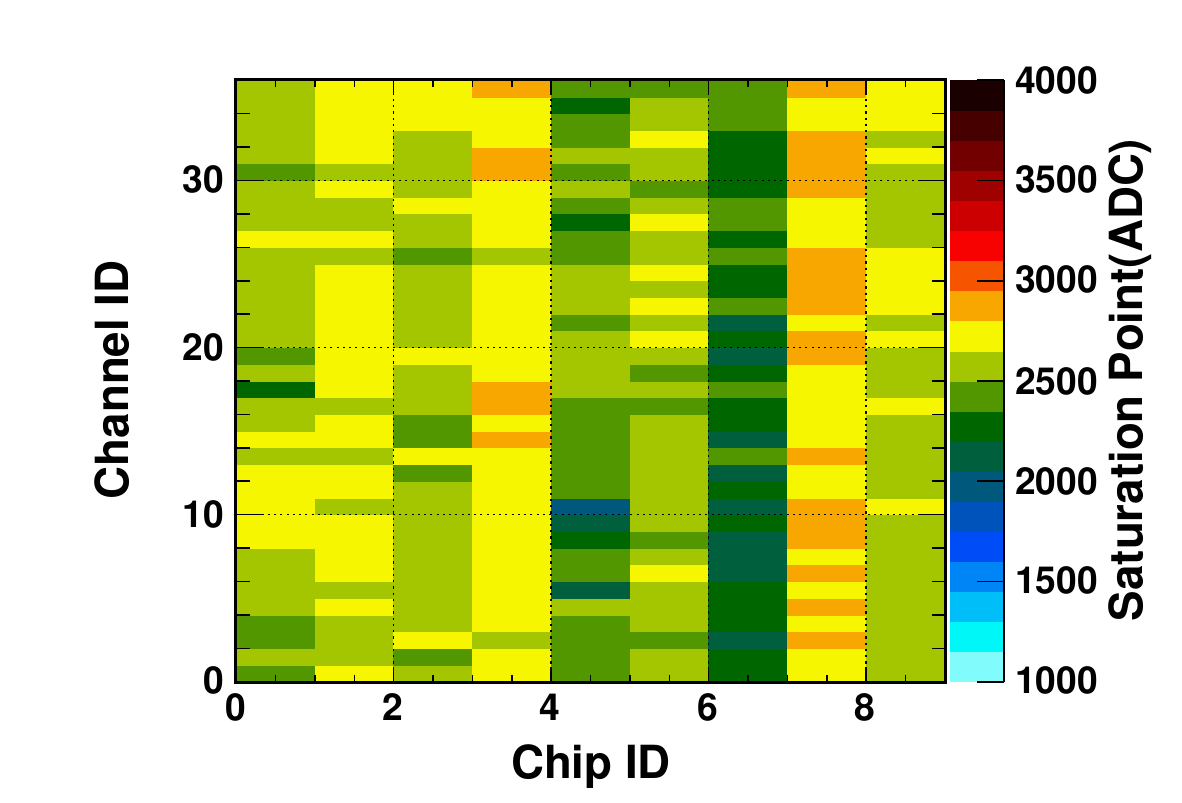}
            \label{Fig:SP_map}
        }
        \caption{Results of gain ratio calibration for AHCAL channels: (a)gain ratio. (b)saturation point. (c) Gain ratios of all sensitive layers. The last two layers exhibited different behavior because they utilize NDL SiPMs. (d)Saturation points varied from chip to chip within the sensitive layer.}
        \label{Fig:GR}
    \end{figure}

    The gain ratios and saturation points of all AHCAL channels were calibrated by the charge injection. The distributions of gain ratios and saturation points were depicted in \cref{Fig:GR}. 

    \cref{Fig:GR_layer} shows that AHCAL channels from most of the sensitive layers have similar gain ratio, except for the last two layers due to the different junction capacitance of NDL SiPMs. 
    \cref{Fig:SP_map} illustrates the differences in saturation points between SPIROC chips, indicating that the saturation points of AHCAL channels are primarily determined by the SPIROC chips. 

\subsection{Cosmic ray test}

\begin{figure}[h]
    \centering
    \includegraphics[width=.35\textwidth]{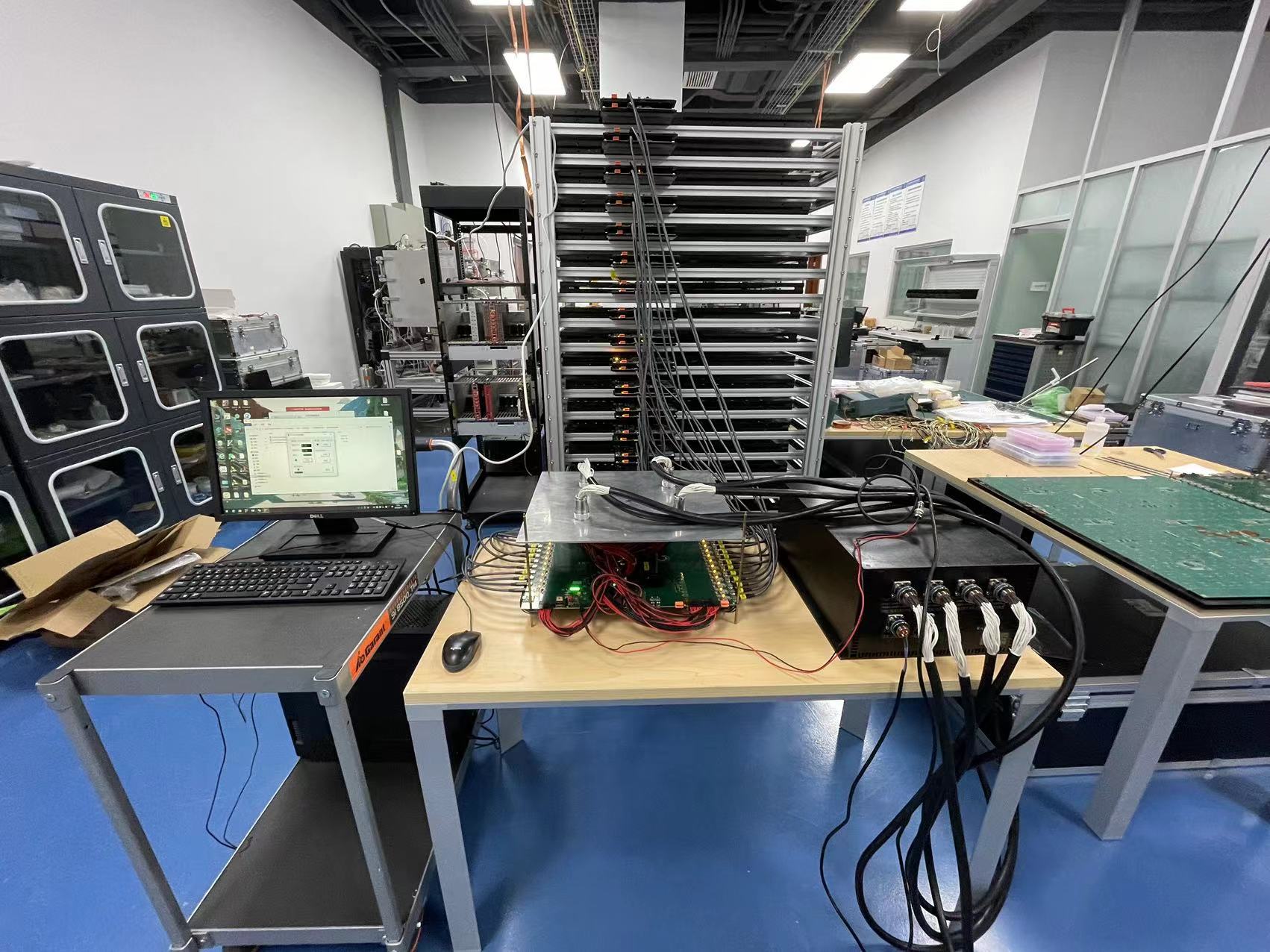}
    \caption{40 sensitive layers were placed in the dedicated support structure to conduct the cosmic ray test. Absorbers were not involved.}
    \label{Fig:cosmic_setup}
\end{figure}

	A cosmic ray test was performed on a dedicated test platform with 40 sensitive layers, as shown in \cref{Fig:cosmic_setup}. During the cosmic ray test, signals from the top and bottom sensitive layers were directed to the TLU for coincidence measurement, with the DAQ board processing the data acquisition accordingly. The cosmic ray test spanned a month, during which half a million events were recorded. 

\begin{figure}[h]
        \centering
        \subfigure[]{
            \centering
            \includegraphics[width=.2\textwidth]{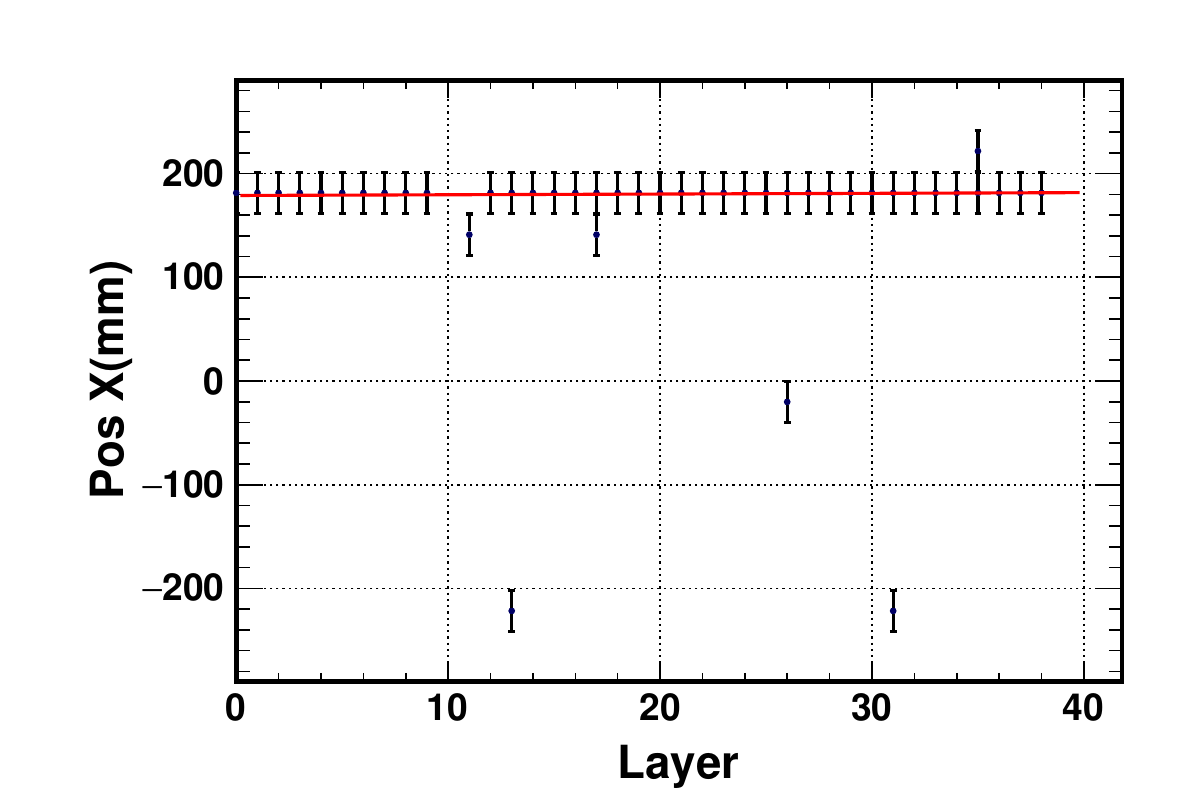}
            \label{Fig:Track_X}
        }
        \subfigure[]{
            \centering
            \includegraphics[width=.2\textwidth]{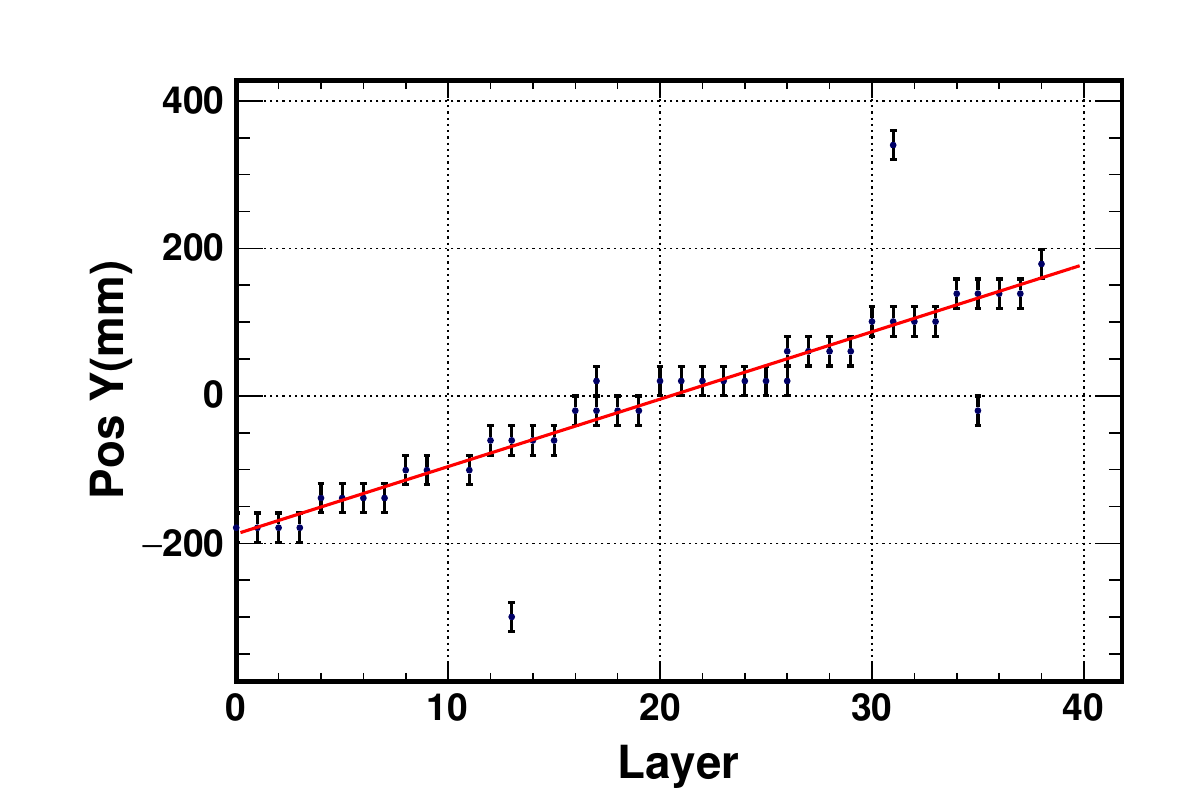}
            \label{Fig:Track_Y}
        }
        \subfigure[]{
            \centering
            \includegraphics[width=.2\textwidth]{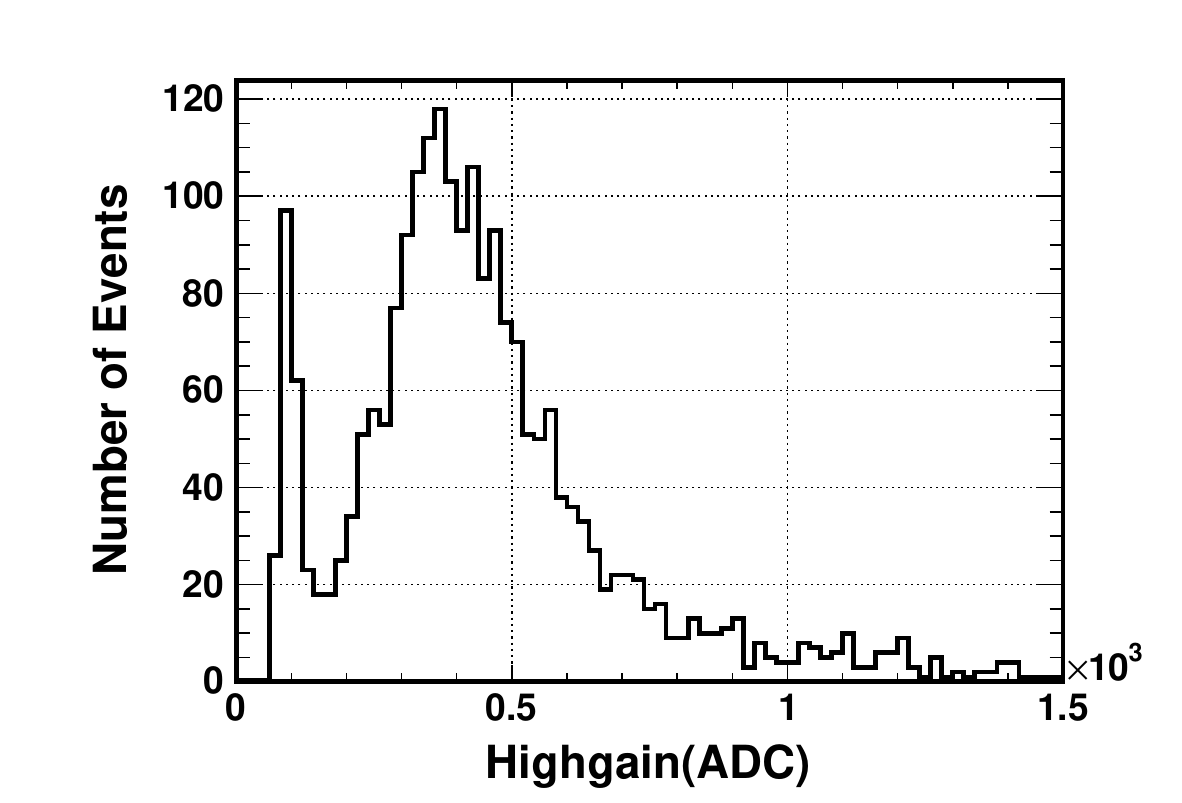}
            \label{Fig:MIP_bc}
        }
        \subfigure[]{
            \centering
            \includegraphics[width=.2\textwidth]{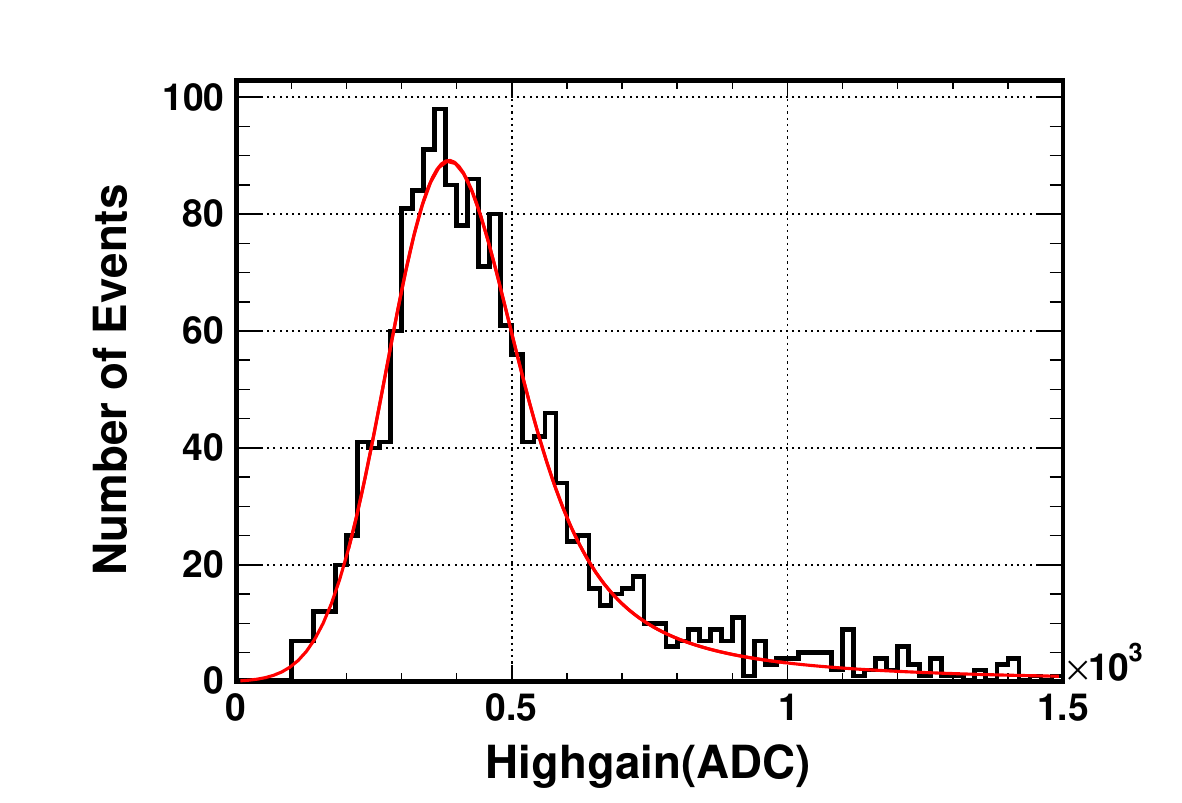}
            \label{Fig:MIP_ac}
        }
        \caption{Displays of cosmic track fittings: (a) X-Z direction, (b) Y-Z direction. Cosmic ray spectra for a chip: (c) before track fitting and noise hit rejection, (d) after track fitting and noise hit rejection.}
        \label{Fig:Track_Fitting}
    \end{figure}

    \cref{Fig:MIP_bc} illustrates the cosmic ray response for AHCAL channels in a chip, showing noticeable dark noise. To suppress this dark noise, a threshold of \SI{100}{ADC} was applied to each hit in the offline analysis, and track fitting was performed. The track fitting was conducted using hits that were the only hit in their respective layers. A minimum of 15 such hits was required in each event. 
  
    \cref{Fig:Track_Fitting} illustrates the fitting of a cosmic event. Some poorly fitted events with high chi-square values were excluded, resulting in approximately 20,000 events displaying good tracking. In these events, noise hits were effectively eliminated by rejecting hits that were more than \SI{120}{mm} away from the fitted track.
    
    \cref{Fig:MIP_ac} demonstrates the impact of the track fitting and noise hit rejection. The MIP peak was fitted using a convolution of Gaussian and Landau functions. After pedestal extraction, the MPV obtained from the fitting was 347 ADC, roughly corresponding to 17 photon electrons according to the LED calibration results.

    \begin{figure}[h]
        \centering
        \includegraphics[width=.35\textwidth]{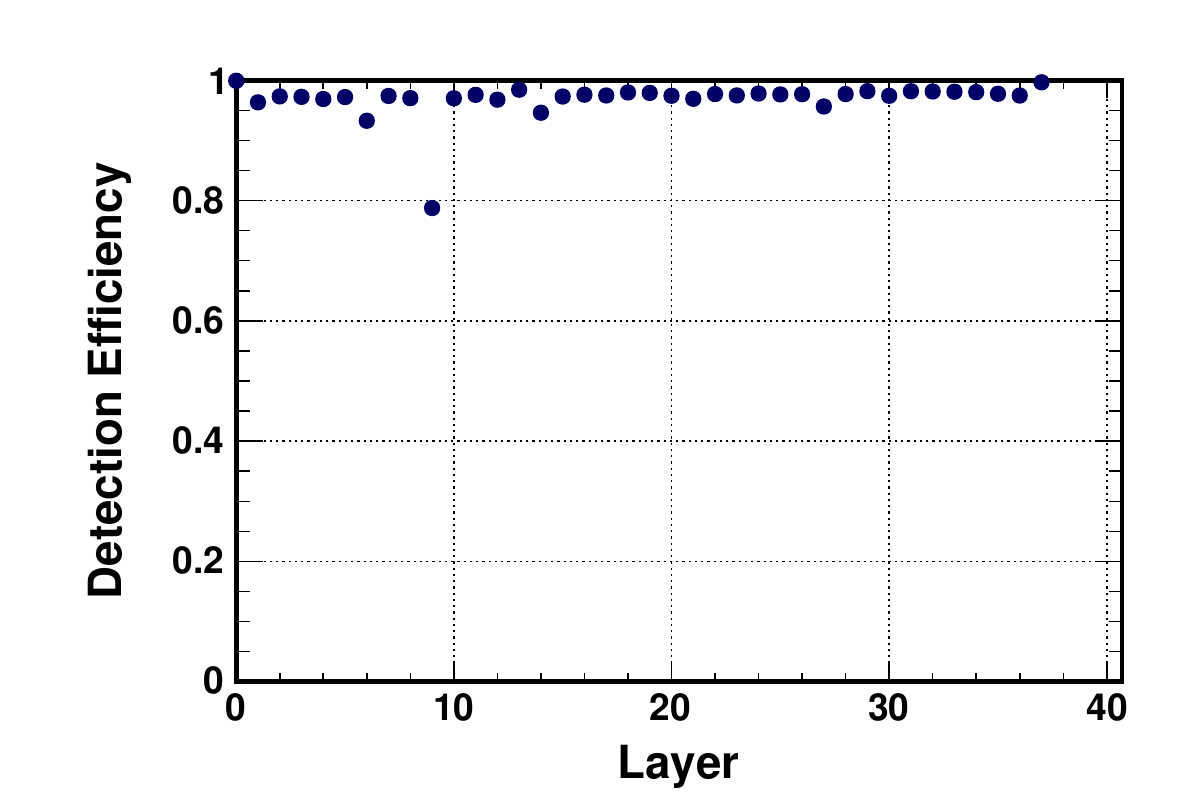}
        \caption{Detection efficiency of AHCAL sensitive layers for cosmic ray events.}
        \label{Fig:cosmic_eff}
    \end{figure}

    It is important to note that the MPV varies from chip to chip due to the differences in SiPM voltages and SPIROC chip configurations. These variations in MPV resulted in differences in detection efficiency between sensitive layers, as shown in \cref{Fig:cosmic_eff}. The majority of AHCAL sensitive layers demonstrated a detection efficiency of approximately 97\%, with approximately 2\% of undetected events attributed to dead areas between scintillator tiles. Layer 9 showed a lower efficiency due to improper configurations, which were subsequently rectified during the beam test.

\section{Summary and outlook}
    Several studies have been conducted on key components of the AHCAL prototype, including the sensitive units, the electronics, and the mechanical structure. These studies optimized the design of the AHCAL, thereby improving the performance of the AHCAL. A 40-layer AHCAL prototype with 12,960 channels was constructed. An electronic test and a cosmic ray test were conducted to the prototype. Pedestal and internal gain of each AHCAL prototype channel were calibrated with the electric test data. The results of the cosmic ray test indicated that the typical MIP response for the AHCAL chip was approximately 17 photoelectrons, while the detection efficiency for cosmic rays in the sensitive layers was approximately 97\%. \textcolor{black}{These results validated the proper functioning of the prototype and confirmed its readiness for the three beam tests conducted between 2022 and 2023.} 
    
    \textcolor{black}{The test beam data of the AHCAL prototype is currently under analysis. Preliminary results indicate that the AHCAL prototype demonstrates excellent imaging capability, enabling detailed visualization of the intricate inner structure of hadron showers. Additionally, its energy resolution reaches approximately $\frac{60\%}{\sqrt{E(\si{GeV})}}$, which satisfies the requirements of the CEPC experiment.} Further results will validate the feasibility of using the AHCAL as a hadron calorimeter for the CEPC experiment and will contribute to enhancements in the CEPC baseline detector design. Additionally, the test beam data presents an excellent opportunity to study hadronic shower shapes and refine the particle flow algorithm.

        
\end{document}